\documentclass[journal]{IEEEtran}

\pdfoutput=1

\usepackage[cmex10]{amsmath}
\usepackage{cite}

\usepackage{latexsym}
\usepackage{graphics}
\usepackage{amssymb}
\usepackage{amsthm}

\usepackage{cite}
\usepackage{array}
\usepackage{mdwmath}
\usepackage{mdwtab}
\usepackage[tight,footnotesize]{subfigure}
\usepackage{graphicx}

\usepackage{stfloats}
\fnbelowfloat
\usepackage{url}

\ifCLASSOPTIONcompsoc
\usepackage[tight,normalsize,sf,SF]{subfigure}
\else
\usepackage[tight,footnotesize]{subfigure}
\fi

\def\tr{\mathrm{tr}}
\def\trace{\mathrm{tr}}

\newcommand{\argmax}[1]{{\underset{{#1}}{\mathrm{arg\,max}}}}

\newcommand{\vect}[1]{\mathbf{#1}}

\newcommand{\maximize}[1]{{\underset{{#1}}{\mathrm{maximize}}}}
\newcommand{\minimize}[1]{{\underset{{#1}}{\mathrm{minimize}}}}

\theoremstyle{remark}
\newtheorem{remark}{Remark}

\newtheorem{theorem}{Theorem}

\newtheorem{lemma}{Lemma}
\newtheorem{definition}{Definition}

\newcommand{\mysmallarraydecl}{\renewcommand{%
\IEEEeqnarraymathstyle}{\scriptscriptstyle}%
\renewcommand{\IEEEeqnarraytextstyle}{\scriptsize}%
\renewcommand{\baselinestretch}{1.1}%
\settowidth{\normalbaselineskip}{\scriptsize
\hspace{\baselinestretch\baselineskip}}%
\setlength{\baselineskip}{\normalbaselineskip}%
\setlength{\jot}{0.25\normalbaselineskip}%
\setlength{\arraycolsep}{2pt}}

\newenvironment{lined}[1]%
 {\begin{center}\begin{minipage}{#1}\hrule\medskip}
 {\vspace{-1ex}\hrule \end{minipage}\end{center}}

\begin{document}

\title{Robust Monotonic Optimization Framework for\\  Multicell MISO Systems}

\author{Emil Bj\"ornson,~\IEEEmembership{Member,~IEEE,}
        Gan Zheng,~\IEEEmembership{Member,~IEEE,}
        Mats Bengtsson,~\IEEEmembership{Senior Member,~IEEE,}
        and\\ Bj\"orn~Ottersten,~\IEEEmembership{Fellow,~IEEE}
        \thanks{\copyright 2012 IEEE. Personal use of this material is permitted. Permission from IEEE must be obtained for all other uses, in any current or future media, including reprinting/republishing this material for advertising or promotional purposes, creating new collective works, for resale or redistribution to servers or lists, or reuse of any copyrighted component of this work in other works. The associate editor coordinating the review of this manuscript and approving it for publication was Prof.~Yimin D.~Zhang. The research leading to these results has received funding from the European Research Council under the European Community's Seventh Framework Programme (FP7/2007-2013) / ERC grant agreement number 228044.}%
\thanks{E.~Bj\"ornson and M.~Bengtsson are with the Signal Processing Laboratory, ACCESS
Linnaeus Center, KTH Royal Institute of Technology, SE-100 44
Stockholm, Sweden (e-mail: emil.bjornson@ee.kth.se; mats.bengtsson@ee.kth.se).}%
\thanks{G.~Zheng is with the Interdisciplinary Centre for Security, Reliability and Trust (SnT), University of Luxembourg,
6 rue Richard Coudenhove-Kalergi, L-1359 Luxembourg-Kirchberg, Luxembourg (email: gan.zheng@uni.lu).}%
\thanks{B.~Ottersten is with the Signal Processing Laboratory, ACCESS
Linnaeus Center, KTH Royal Institute of Technology, SE-100 44
Stockholm, Sweden. He is also with the Interdisciplinary Centre for Security, Reliability and Trust (SnT), University of Luxembourg,
6 rue Richard Coudenhove-Kalergi, L-1359 Luxembourg-Kirchberg, Luxembourg (email: bjorn.ottersten@ee.kth.se).}%
\thanks{Digital Object Identifier 10.1109/TSP.2012.2184099}%
}

\markboth{IEEE TRANSACTIONS ON SIGNAL PROCESSING, VOL.~60, NO.~5, MAY 2012}%
{Bj\"ornson \MakeLowercase{\textit{et al.}}: ROBUST MONOTONIC OPTIMIZATION FRAMEWORK FOR  MULTICELL MISO SYSTEMS}

\maketitle

\begin{abstract}
The performance of multiuser systems is both difficult to measure fairly and to optimize. Most resource allocation problems are non-convex and NP-hard, even under simplifying assumptions such as perfect channel knowledge, homogeneous channel properties among users, and simple power constraints. We establish a general optimization framework that systematically solves these problems to global optimality. The proposed \emph{branch-reduce-and-bound (BRB) algorithm} handles general multicell downlink systems with single-antenna users, multiantenna transmitters, arbitrary quadratic power constraints, and robustness to channel uncertainty.
A \emph{robust fairness-profile optimization} (RFO) problem is solved at each iteration, which is a quasi-convex problem and a novel generalization of max-min fairness. The BRB algorithm is computationally costly, but it shows better convergence than the previously proposed outer polyblock approximation algorithm. Our framework is suitable for computing benchmarks in general multicell systems with or without channel uncertainty. We illustrate this by deriving and evaluating a zero-forcing solution to the general problem.
\end{abstract}

\begin{IEEEkeywords}
Branch-reduce-and-bound, dynamic cooperation clusters, fairness-profile, Network MIMO, optimal resource allocation, performance region, worst-case robustness.
\end{IEEEkeywords}

\IEEEpeerreviewmaketitle

\section{Introduction}

\IEEEPARstart{R}{esource} allocation is generally very difficult in multiantenna systems. First of all, it is non-obvious how to measure multiuser system performance. In information theory, the sum capacity provides the highest reliable data throughput \cite{Viswanath2003a}, regardless of the computational complexity and delay resilience required for implementation. Signal processing measures such as the mean squared error (MSE) are, on the other hand, only vaguely connected to the user-experienced service quality \cite{Dohler2011a}. Secondly, multiuser systems are limited by interference, requiring considerations between optimizing total performance and guaranteeing individual user service. Cellular users are often highly heterogeneous, both in average channel gain and delay sensitivity \cite{Huh2010a}, making it tricky to even define fairness among users. Thirdly, simplifying assumptions on channel state information (CSI), power constraints, synchronization, and performance measures are required to achieve tractable mathematical problems.

A key to efficient performance optimization is to formulate it as a convex problem, making the global solution achievable through efficient algorithms \cite{SDPT3}. Convex formulations for downlink transmission were developed in \cite{Bengtsson2001a} and gradually extended in \cite{Bjornson2011a, Wiesel2006a,Yu2007a} to general power constraints and multicell conditions. Efficient algorithms, based on fixed point iterations, were developed in \cite{Schubert2004a,Schubert2005a}. The convexity was achieved by assuming perfect CSI and pre-defined user performance constraints, thus ignoring how to select these optimally. The extension to maximizing the worst performance among all users is achieved by solving a series of these convex problems
\cite{Wiesel2006a,Yu2007a,Schubert2004a,Schubert2005a}. The requirement of perfect CSI can also be relaxed using robust optimization techniques \cite{Bental2009a}. By assuming ellipsoidal uncertainty regions, convex formulation to the aforementioned problems can be achieved under worst-case robustness \cite{Zheng2008b,Vucic2009a,Shenouda2009a,Vucic2009b,Tajer2011a}. In particular, \cite{Shenouda2009a,Vucic2009b,Tajer2011a} discuss such robustness in a few special multicell scenarios. Robustness can also be defined probabilistically (i.e., with outage probabilities), but (conservative) bounds and approximations are required to achieve convex formulations in these cases \cite{Chalise2007a,Shenouda2008a,Wang2011a}.

Based on the references above, the multicell resource allocation can be solved in polynomial time either under fixed user performance constraints or if the goal is to maximize the worst (i.e., max-min) performance among users. Under general system performance measures, the global solution cannot be achieved efficiently;
 \cite{Liu2011a} shows that sum performance, proportional fairness, and harmonic mean optimizations are all NP-hard problems. However, such problems can still be solved with global convergence and optimality using the framework of monotonic optimization, developed in \cite{Tuy2000a,Tuy2005a}. The outer polyblock approximation is an algorithm in this framework \cite{Tuy2000a}, and applications to single-cell \cite{Brehmer2009b,Brehmer2009a} and multi-cell transmission \cite{Jorswieck2010a,Brehmer2010a,Utschick2012a} with perfect CSI have appeared in literature. Unfortunately, the polyblock algorithm requires a very large number of iterations to achieve accurate results, thus limiting usage to systems with no more than a handful of users \cite{Tuy2005a}.

In this paper, we propose a \emph{robust monotonic optimization framework} for general multicell scenarios with imperfect CSI. The framework can be applied for any system performance measure that increases monotonically in the performance of each user, which of course is satisfied by all reasonable measures. Convergence to the global optimum is achieved through a \emph{branch-reduce-and-bound (BRB) algorithm} that builds upon previous work in \cite{Tuy2005a}, and we show far better convergence than the polyblock algorithm in \cite{Brehmer2009a,Jorswieck2010a,Brehmer2010a,Utschick2012a}. Each iteration of the BRB algorithm solves a quasi-convex subproblem. It is called a \emph{robust fairness-profile optimization}, meaning that each user has a constraint on the lowest acceptable performance level and attains a predefined percentage of all performance above these levels. We show how to formulate this problem efficiently under worst-case robustness, extending results in \cite{Zhang2010a, Karipidis2010a} for perfect CSI. Observe that the BRB algorithm solves a high-complexity (NP-hard) problem and is therefore mainly useful as a benchmark in system level evaluations of suboptimal low-complexity algorithms, although good lower bounds on the optimal solution is achieved in a few iterations.

The structure and contributions of the paper are:

\begin{itemize}
\item The general multicell system model of \cite{Bjornson2011a,Bjornson2010d} with dynamic cooperation clusters is introduced in Section \ref{section_system_model}. User performance is measured by arbitrary monotonic functions of the worst-case MSE and system performance is an arbitrary monotonic function of each user's performance. The concept of a robust performance region is defined and important properties are proved.

\item In Section \ref{section_opt_fairnessprofile}, \emph{robust fairness-profile optimization} (RFO) is introduced as a novel extension to standard max-min performance optimization problems. This problem is shown to be quasi-convex under worst-case robustness and a simple solution algorithm is given.

\item In Section \ref{section_monotonic_optimization}, a novel framework for solving general \emph{robust monotonic optimization problems} is proposed. Convergence to the global optimum of this NP-hard problem is achieved by a branch-reduce-and-bound (BRB) algorithm over the robust performance region.

\item To find initial performance bounds and illustrate the benchmarking capability of the BRB algorithm, Section \ref{section_low-complexity_strategies} derives an approximation of the general optimization problem. By adding interference constraints and pretending to have perfect CSI, a convex formulation is achieved.

\item The proposed framework is evaluated numerically in Section \ref{section_numerical_examples}. The robust performance region is illustrated and the strategy of Section \ref{section_low-complexity_strategies} compared with the global optimum. The computational complexity of RFO is shown to be manageable and the BRB algorithm shows better convergence than the polyblock algorithm in \cite{Brehmer2010a,Utschick2012a}.
\end{itemize}

We have previously applied this framework to the different problem of robust coordinated beamforming with perfect intracell CSI and uncertain intercell CSI; see \cite{Bjornson2011d}.
The previous paper maximized functions of the signal-to-interference-and-noise ratios (SINRs), instead of functions of the MSEs.

\subsection{Notation}

Boldface (lower case) is used for column
vectors, $\vect{x}$, and (upper case) for matrices, $\vect{X}$.
Let $\vect{X}^T$, $\vect{X}^H$, and $\vect{X}^*$ denote the
transpose, the conjugate transpose, and the conjugate of
$\vect{X}$, respectively. For Hermitian square matrices $\vect{X},\vect{Y}$, $\vect{X} \succ \vect{Y}$ and $\vect{X} \succeq \vect{Y}$ means that
$\vect{X}-\vect{Y}$ is positive definite and semi-definite, respectively. $\vect{I}_M,\vect{0}_M \in \mathbb{R}^{M \times M}$ denote identity and zero matrices, respectively. The $L_i$-norm of $\vect{x}$ is $\|\vect{x}\|_i$. $\vect{1}_M \in \mathbb{R}^{M \times 1}$ is a vector with ones. The set of non-negative real $n$-dimensional vectors is denoted $\mathbb{R}_+^{n}$. Element-wise (strict) inequality for vectors $\vect{x},\vect{y}$ is denoted $\vect{x} \leq \vect{y}$ ($\vect{x} < \vect{y}$).

\section{System Model \& Performance Measures}
\label{section_system_model}

We consider a multiple-input-single-output (MISO) system with $K_t$ transmitting base stations and $K_r$ receiving users. The $j$th base station is denoted
$\textrm{BS}_j$ and has $N_j$ antennas. The total number of transmit antennas is $N=\sum_{j=1}^{K_t} N_j$. The $k$th user is denoted $\textrm{MS}_k$, has a single (effective) antenna\footnote{This model also applies to simple multi-antenna receivers that fix a receive beamformer (e.g., antenna selection) prior to transmission
optimization.}, and is a simple receiver:

\begin{definition} \label{def_simple_receiver}
A \emph{simple receiver} decodes its designated signal
\begin{itemize}
\item As consisting of a scalar-coded data symbol $s_k$ multiplied with a transmit beamforming vector $\vect{v}_k \in \mathbb{C}^{N \times 1}$;
\item While treating co-user interference as noise (i.e., without trying to decode and subtract interfering signals).
\end{itemize}
\end{definition}

Under these conditions, the transmission should obviously satisfy the first property.
The use of transmit beamforming is actually optimal under single-user detection (i.e., the second property) if perfect CSI is available \cite{Mochaourab2011a,Shang2010a,Bjornson2011a}, while \cite{Song2011a} provides conditions on its optimality under channel uncertainty. From an information theoretic perspective, transmit beamforming and simple receivers are suboptimal \cite{Shang2009b} but these assumptions are of practical importance to achieve low-complexity receivers and power efficiency.

In a general multicell scenario, some users are served in a coordinated manner by multiple transmitters. In addition, some transmitters and
receivers are very far apart, making it impractical to estimate and
separate the interference on these channels from the noise. To
capture these properties, we apply the dynamic coordination
framework of \cite{Bjornson2011a,Bjornson2010d}:

\begin{definition} \label{def_dynamic_clusters}
\emph{Dynamic cooperation clusters} means that $\textrm{BS}_j$
\begin{itemize}
\item Has channel estimates to receivers in $\mathcal{C}_j
\subseteq \{1,\!...,\!K_r\!\}$, while the interference generated to receivers $\bar{k} \not \in \mathcal{C}_j$ is treated as part of the background noise;
\item Serves the receivers in $\mathcal{D}_j \subseteq
\mathcal{C}_j$ with data.
\end{itemize}
\end{definition}

\begin{figure}[t!]
\begin{center}
\includegraphics[width=\columnwidth]{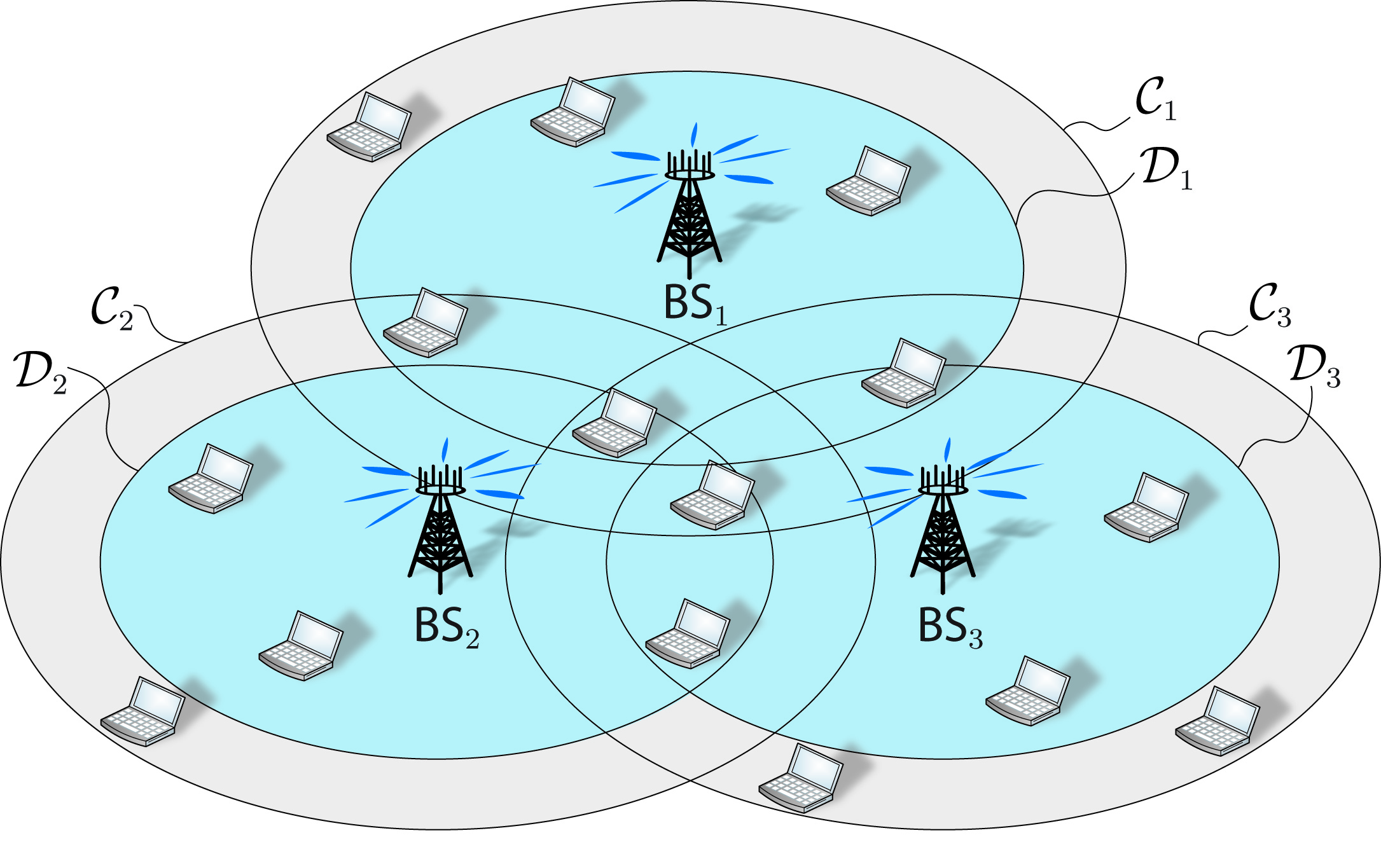}
\end{center} \vskip -5mm
\caption{Schematic intersection between three cells. $\textrm{BS}_j$
serves users in the inner circle ($\mathcal{D}_j$), while it coordinates interference to  users in the outer circle ($\mathcal{C}_j$). Ideally, negligible interference is caused to users outside both circles.}\label{figure_system_model} \vskip -3mm
\end{figure}

This coordination framework is characterized by the sets $\mathcal{C}_j,\mathcal{D}_j$, and the mnemonic rule is that $\mathcal{D}_j$ describes \emph{data} from transmitter $j$ while $\mathcal{C}_j$ describes \emph{coordination} from transmitter $j$. To reduce backhaul signaling of data, the cardinality of
$\mathcal{D}_j$ is typically smaller than that of $\mathcal{C}_j$. These
sets are illustrated in Fig.~\ref{figure_system_model} and are selected based on long-term channel gains (see \cite{Bjornson2011a} for details).
To enable coordinated transmissions, perfect phase coherence and synchronous interference is assumed between transmitters that serve users jointly (see \cite{Zhang2008a}).

The narrowband, flat-fading channel from $\textrm{BS}_j$ to $\textrm{MS}_k$ is $\vect{h}_{jk} \in \mathbb{C}^{N_j \times 1}$. The combined channel from all transmitters
is denoted $\vect{h}_{k}=[\vect{h}_{1k}^T \ldots \vect{h}_{K_t k}^T
]^T \in \mathbb{C}^{N \times 1}$. The received signal at $\textrm{MS}_k$ is modeled as
\begin{equation} \label{eq_system_model}
y_{k}= \vect{h}_{k}^H \vect{C}_k \sum_{\bar{k}=1}^{K_r}
\vect{D}_{\bar{k}} \vect{v}_{\bar{k}} s_{\bar{k}} + n_{k}
\end{equation}
where the scalar-coded data symbol $s_{k}$ for $\textrm{MS}_k$ is assumed to be zero-mean and have unit-variance (without loss of generality). The block-diagonal matrix $\vect{D}_k \in
\mathbb{C}^{N \times N}$ selects the transmit antennas that send $s_{k}$ and is defined as
\begin{equation}
\begin{split}
[\vect{D}_{k}]_{\textrm{diagonal block } j} =
\begin{cases}
\vect{I}_{N_j}, & \textrm{if } k \in \mathcal{D}_j,\\
\vect{0}_{N_j}, & \textrm{if } k \not \in \mathcal{D}_j.
\end{cases}
\end{split}
\end{equation}
Observe that $\vect{D}_k \vect{v}_k$ is the effective beamforming vector, but we will optimize over $\vect{v}_k$ for notational convenience; any (reasonable) solution to the optimization problems herein will satisfy $\vect{v}_k = \vect{D}_k \vect{v}_k$.

Similarly, $\vect{C}_k \in \mathbb{C}^{N \times N}$ selects signals from transmitters that have channel estimates to $\textrm{MS}_k$ (i.e., those with non-negligible channels). This block-diagonal matrix is defined as
\begin{equation}
\begin{split}
[\vect{C}_{k}]_{\textrm{diagonal block } j} =
\begin{cases}
\vect{I}_{N_j}, & \textrm{if } k \in \mathcal{C}_j,\\
\vect{0}_{N_j}, & \textrm{if } k \not \in \mathcal{C}_j.
\end{cases}
\end{split}
\end{equation}
The noise and remaining (weak) interference are given by the circular-symmetric complex Gaussian term $n_k \in \mathcal{CN}(0,\sigma_k^2)$.

The transmission (i.e., selection of beamforming vectors) is limited by $L$ quadratic power constraints
\begin{equation} \label{eq_power_constraints}
\sum_{k=1}^{K_r} \vect{v}_k^H \vect{Q}_l \vect{v}_k \leq q_l  \quad l=1,\ldots,L,
\end{equation}
where $\vect{Q}_l \in \mathbb{C}^{N \times N}$ are Hermitian positive semi-definite matrices for all $l$. To make sure that the power is constrained in all spatial dimensions, these matrices satisfy $\sum_{l=1}^L \vect{Q}_l \succ \vect{0}_N$.

\subsection{Channel State Information and Robustness}

In practice, transmitters have uncertain CSI. The uncertainty originates from a variety of sources, including
channel estimation, feedback quantization, hardware deficiencies, and delays in CSI acquisition on fading channels.
It is common to assume an additive error model \cite{Bental2009a,Zheng2008b,Vucic2009a,Vucic2009b,Shenouda2009a,Tajer2011a,Chalise2007a,Shenouda2008a,Wang2011a} with
\begin{equation} \label{eq_additive_error}
\vect{h}_k = \widehat{\vect{h}}_k + \boldsymbol{\epsilon}_k \quad \forall k
\end{equation}
where $\widehat{\vect{h}}_k=[\widehat{\vect{h}}_{1 k}^T \ldots \widehat{\vect{h}}_{K_t k}^T]^T \in \mathbb{C}^{N \times 1}$ is the uncertain CSI of the combined channel vector $\vect{h}_k$ and $\boldsymbol{\epsilon}_k \in \mathbb{C}^{N \times 1}$ is the combined error vector.
This model can, for instance, be motivated by viewing channel estimation as the main source of uncertainty \cite{Bjornson2010a}.\footnote{Under training-based MMSE channel estimation \cite{Bjornson2010a}, the error takes the form of \eqref{eq_additive_error}. The stochastic error vector is $\boldsymbol{\epsilon}_k \in \mathcal{CN} (\vect{0},\vect{E}_k)$ under Rayleigh fading. If the channel from each base station to user $k$ is estimated separately, then the estimation error covariance matrix $\vect{E}_k$ becomes block-diagonal.\label{footnote}}
Observe that both the channel estimate and the error should be set to zero for all $\vect{h}_{jk}$ with $k \not \in \mathcal{C}_j$.

The stochastic distribution for $\boldsymbol{\epsilon}_k$ is unbounded\footnote{This holds for Rayleigh fading channels, while practical estimation errors of course are bounded but can be very large.}, thus communication cannot be robust towards any error. This is usually handled by only considering a subset of error vectors, \emph{the uncertainty set}, that has high probability \cite{Bental2009a,Zheng2008b,Vucic2009a,Vucic2009b,Shenouda2009a,Tajer2011a,Chalise2007a,Shenouda2008a,Wang2011a}. If this set is included in the resource allocation (i.e., optimization with acceptable outage probability), approximations are required to achieve tractable problem formulations \cite{Chalise2007a,Shenouda2008a,Wang2011a}. Herein, we consider a fixed uncertainty set and maximize the worst-case performance over this set \cite{Shenouda2009a,Vucic2009b,Tajer2011a}. This approach is convenient as it can provide convex problem formulations, but is often accused of giving conservative performance results \cite{Gershman2010a}. However, this is the result of using ill-structured uncertainty sets and can be avoided by proper selection of these sets.\footnote{In the probabilistic approach, the guaranteed performance is maximized under a given outage probability. Using an optimal precoding solution, one can create a set $\mathcal{U}$ of all error vectors that gives exactly the optimal guaranteed performance (or better). If $\mathcal{U}$ is used as the uncertainty set in the worst-case approach, it will provide the same optimal precoding solution.}

For analytical convenience and motivated by channel estimation\footnote{Continuing the estimation example in a previous footnote, recall that $\boldsymbol{\epsilon}_k
\in \mathcal{CN}(\vect{0},\vect{E}_{k})$. Thus, $\boldsymbol{\epsilon}_k$ belongs with probability $\rho$ to the ellipsoidal set $\{ \boldsymbol{\epsilon}_k : 2 \boldsymbol{\epsilon}_k^H \vect{E}_k^{-1} \boldsymbol{\epsilon}_k \leq \chi_{\rho}^{2}(2N)\}$, where $\chi_{\rho}^{2}(n)$ is the $\rho$-percentile of the $\chi^{2}(n)$-distribution. If we limit the robustness to this set, the channel uncertainty is given by \eqref{eq_def_uncertainty_sets} using $\vect{B}_{k} = \sqrt{ \frac{\chi_{\rho}^{2}(2N)}{2} } \vect{E}_k^{1/2}$. To enforce higher or lower robustness to errors on channels from some base stations, one can use different weights on the diagonal blocks of $\vect{B}_{k}$.} \cite{Vucic2009b,Shenouda2009a,Tajer2011a}, we concentrate on (compact) ellipsoidal channel uncertainty sets
\begin{equation} \label{eq_def_uncertainty_sets}
\mathcal{U}_k(\widehat{\vect{h}}_k,\vect{B}_{k}) = \left\{ \vect{h}_k : \vect{h}_k = \widehat{\vect{h}}_k + \vect{B}_{k} \tilde{\boldsymbol{\epsilon}}_{k}, \,\, \| \tilde{\boldsymbol{\epsilon}}_{k} \|_2 \leq 1  \right\}
\end{equation}
where $\vect{B}_{k} \in \mathbb{C}^{N \times N}$ defines the shape of the ellipsoid. Since many uncertainty sources are independent between transmitters
(e.g., estimation and quantization are done separately), $\vect{B}_{k}$ is typically block-diagonal in multicell systems. However, the analysis herein is not limited to such $\vect{B}_{k}$. Other types of compact uncertainty sets (including separate sets for each $\vect{h}_{jk}$ and probabilistic robustness) are discussed in Section \ref{subsection_extensions}.

While $\mathcal{U}_1,\ldots,\mathcal{U}_{K_r}$ represent the CSI at the transmitter side, each $\text{MS}_k$ is assumed to only have a local estimate of $\vect{h}_{k}$. Thus, the receivers are unaware of co-user interference and precoding vectors, and are therefore assumed to be optimized by the transmitters and told how to process their received signals. Observe that the performance can be improved by, for example, estimating the optimal equalizers at each receiver based on the effective channels with precoding, but this requires additional training overhead that might be unavailable.

\subsection{Examples: Two Simple Multicell Scenarios}

The purpose of the above system model is to jointly describe and analyze a variety of multicell scenarios. Typical examples are ideal network MIMO\footnote{Ideal multiuser coordination is commonly called network multiple-input multiple-output (MIMO), even in the case of single-antenna users.} \cite{Karakayali2006a} (where all transmitters serve all users) and MISO interference channels \cite{Jorswieck2008b,Shang2010a} (with only one unique user per transmitter):

\subsubsection{Ideal Network MIMO} All transmitters serve and coordinate interference to all users, meaning that $\vect{D}_k=\vect{C}_k=\vect{I}_N$ for all $k$. If a total power constraint is used, then $L=1$ and $\vect{Q}_1=\vect{I}_N$. If per-antenna constraint are used, then $L=N$ and $\vect{Q}_l$ is only non-zero at the $l$th diagonal element. If perfect CSI is available, then $\vect{B}_k=\vect{0}_N$ and thus $\mathcal{U}_k = \{ \widehat{\vect{h}}_k \}$ for all $k$.

\subsubsection{Two-user MISO Interference Channel} Let $\textrm{BS}_k$ serve $\textrm{MS}_k$ and coordinate interference to the other receiver. Then, $\vect{D}_1=\left[\begin{IEEEeqnarraybox*}[\mysmallarraydecl]
[c]{,c,c,}
\vect{I}_{N_1}& \vect{0}\\
\vect{0}& \vect{0}%
\end{IEEEeqnarraybox*}\right]$ and $\vect{D}_2=\left[\begin{IEEEeqnarraybox*}[\mysmallarraydecl]
[c]{,c,c,}
\vect{0} & \vect{0}\\
\vect{0}& \vect{I}_{N_2}%
\end{IEEEeqnarraybox*}\right]$, while $\vect{C}_1=\vect{C}_2=\vect{I}_N$. If each transmitter has its own total power constraint, then $L=2$ and $\vect{Q}_l=\vect{D}_l$ for $l=1,2$. If each transmitter estimates its channel independently, then a block-diagonal matrix $\vect{B}_k=\left[\begin{IEEEeqnarraybox*}[\mysmallarraydecl]
[c]{,c,c,}
\vect{B}_{k1}& \vect{0}\\
\vect{0}& \vect{B}_{k2}%
\end{IEEEeqnarraybox*}\right]$ is used to define the uncertainty sets $\mathcal{U}_k$. If channel estimation is the main source of uncertainty, then $\vect{B}_{kj}$ is a scaled version of the estimation error variance for $\vect{h}_{jk}$ \cite{Bjornson2010a}. The scaling decides the amount of error that the system is robust to.

\subsection{User Performance}

\begin{figure}[t!]
\begin{center}
\includegraphics[width=\columnwidth]{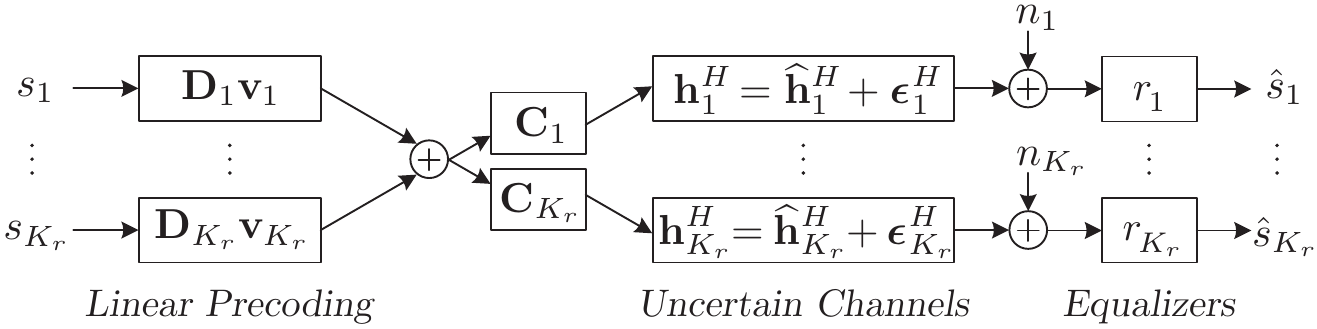}
\end{center} \vskip -5mm
\caption{Block diagram of the downlink multicell system.
Linear precoding is applied to each data stream and $\vect{D}_k$ decides which antennas that can transmit to user $k$.
The channel uncertainty is modeled by additive errors $\boldsymbol{\epsilon}_k$, while $\vect{C}_k$ removes negligible channels that are included in the additive noise $n_k$. User $k$ applies the equalizing coefficient $r_k$ to estimate its data signal.}\label{figure_block_model} \vskip -3mm
\end{figure}

The user performance is based on the MSE. $\text{MS}_k$ uses an equalizing coefficient $r_k$ to achieve an estimate $\hat{s}_k = r_k y_{k}$ of the transmitted data signal $s_k$. Thus, the MSE in the data estimation at $\text{MS}_k$ is $\text{MSE}_k = \mathbb{E} \{ |\hat{s}_k - s_k|^2 \}$ and becomes
\begin{equation} \label{eq_MSEk}
\begin{split}
&\text{MSE}_k = \| r_k \vect{h}_{k}^H \vect{C}_k [\vect{D}_{1} \vect{v}_1 \,\ldots\,  \vect{D}_{K_r} \vect{v}_{K_r}] -  \vect{e}_k^T   \|_2^2 + |r_k|^2 \sigma_k^2 \\
& \,\,\, = \underbrace{|r_k \vect{h}_{k}^H \vect{C}_k \vect{D}_{k} \vect{v}_k - 1|^2}_{\text{signal distortion}} + \underbrace{\sum_{\bar{k} \neq k} |r_k \vect{h}_{k}^H \vect{C}_k \vect{D}_{\bar{k}} \vect{v}_{\bar{k}} |^2}_{\text{co-user interference}} + \underbrace{|r_k|^2 \sigma_k^2}_{\text{noise}}
\end{split}
\end{equation}
where $\vect{e}_k$ denotes the $k$th column of $\vect{I}_{K_r}$. For MSE optimization, it suffices to consider real-valued $r_k\geq0$ as any complex phase can be included in the beamforming vector $\vect{v}_k$ without affecting the MSE in \eqref{eq_MSEk}. A block diagram of the system model is shown in Fig.~\ref{figure_block_model}.

Since the MSE describes the average squared distance between $s_k$ and its estimate $\hat{s}_k$, it should be small. The range of reasonable\footnote{We can always disregard the received signal by setting $r_k=0$ and achieve $\text{MSE}_k=1$, thus $\text{MSE}_k>1$ is always suboptimal.} MSE values is
\begin{equation} \label{eq_MSE_range}
0 < \text{MSE}_k \leq \mathbb{E} \{ |s_k|^2  \} = 1
\end{equation}
where the lower bound assumes negligible noise and interference, while the upper bound is the original signal variance.

Herein, the performance of $\textrm{MS}_k$ is measured by a continuous function $g_k(\textrm{MSE}_{k})$ of the MSE. Our convention is that good performance means large positive values, thus $g_k(\cdot)$ is a
\emph{strictly decreasing}\footnote{A function $g: \mathbb{R}_+ \rightarrow \mathbb{R}$ is \emph{strictly decreasing} if for any $x,x'\in \mathbb{R}_+$ such that
$x > x'$ it follows that $g(x)<g(x')$.} function. From \eqref{eq_MSE_range}, the function is bounded as
\begin{equation}
0 = g_k(1)  \leq g_k(\textrm{MSE}_{k}) < g_k(0)
\end{equation}
where we assumed $g_k(1)=0$ for notational convenience. Performance measures that can be expressed in this way are, for instance, bit error rate (BER), data rate, SINR, and the MSE itself. If the equalizing coefficients $r_k$ are based on perfect CSI, there are simple expressions for these utilities \cite{Palomar2003a}. CSI uncertainty makes it hard to derive closed-form expressions, but a simple relationship is given in \cite[Lemma 1]{Shenouda2009a}.

The user performance is limited by the power constraints in \eqref{eq_power_constraints}, but also by co-user interference. The MSE in \eqref{eq_MSEk} improves if the interference is decreased, but this will degrade the MSEs for other users. Under worst-case robustness, this relationship is characterized by the robust performance region:

\begin{definition}
The \emph{robust performance region} $\mathcal{R} \subset
\mathbb{R}_+^{K_r}$ is
\begin{equation} \label{eq_performance_region}
\begin{split}
\mathcal{R} =
\Big\{ \big( g_{1}(\widetilde{\textrm{MSE}}_1), & \ldots,
g_{K_r}(\widetilde{\textrm{MSE}}_{K_r}) \big) : \\ &
 ( \vect{v}_1,\ldots,\vect{v}_{K_r} ) \in
\mathcal{V}, \quad r_k\geq 0 \,\, \forall k
\Big\}
\end{split}
\end{equation}
where the worst-case MSE is denoted
\begin{equation}
\widetilde{\textrm{MSE}}_{k}= \min\left( \max_{\vect{h}_k \in \mathcal{U}_k} \textrm{MSE}_{k}, 1 \right)
\end{equation}
and $\mathcal{V}$ is the set of feasible transmit strategies:
\begin{equation} \label{eq_feasible_transmit_strategies}
\mathcal{V} = \left\{ ( \vect{v}_1,\ldots,\vect{v}_{K_r}  ) : \,
 \sum_{k} \vect{v}_k^H
\vect{Q}_l \vect{v}_k \leq q_l \,\,\, \forall l \right\}.
\end{equation}
\end{definition}

This region describes the performance that can be guaranteed to be simultaneously
achieved by the users. The shape of the $K_r$-dimensional region depends strongly on the effective channels, uncertainty sets, power constraints, and dynamic cooperation clusters. In general, it is a non-convex set, but it can be characterized as normal \cite{Tuy2000a}:
\begin{definition} \label{def_normal}
A set $\mathcal{T} \subset
\mathbb{R}_+^{n}$ is called \emph{normal} if for any point $\vect{x} \in \mathcal{T}$, all $\vect{x}' \in \mathbb{R}_+^{n}$ with  $\vect{x}' \leq \vect{x}$ also satisfy $\vect{x}' \in \mathcal{T}$.
\end{definition}

\begin{lemma} \label{lemma_performance_region_normal}
The robust performance region $\mathcal{R}$ with compact uncertainty sets $\mathcal{U}_1,\ldots,\mathcal{U}_{K_r}$ is a compact and normal set.
\end{lemma}
\begin{IEEEproof}
The proof is given in Appendix \ref{appendix_performance_region_normal}.
\end{IEEEproof}
This means that for any point $\vect{x} \in \mathcal{R}$, all points that give weaker performance than $\vect{x}$ are also in $\mathcal{R}$. This simplifies the search for points in $\mathcal{R}$ that yield good performance; they all lie on the upper boundary $\partial^+ \mathcal{R}$ and this boundary is easy to identify since there are no holes in $\mathcal{R}$.

\begin{definition} \label{def_upper_boundary}
A point $\vect{y}$ is called an \emph{upper boundary point} of a compact normal set $\mathcal{T}$, if $\vect{y} \in \mathcal{T}$ while $\{\vect{y}' \in \mathbb{R}_+^{n}: \, \vect{y}'>\vect{y} \} \cap \mathcal{T} = \emptyset$. The set of all upper boundary points is called the \emph{upper boundary of} $\mathcal{T}$ and is denoted $\partial^+ \mathcal{T}$.
\end{definition}

To determine which point on $\partial^+ \mathcal{R}$ that is preferable, we need a system performance perspective.

\subsection{System Performance}
\label{subsection_system_performance_measures}

While the achievable user performance is represented by the multi-dimensional robust performance region $\mathcal{R}$, the system performance is given by a function $f: \mathcal{R} \rightarrow \mathbb{R}$ that takes a point in $\mathcal{R}$ as input and produces a scalar value. For a given point $\vect{g}=(g_1,\ldots,g_{K_r}) \in \mathcal{R}$, typical examples are
\begin{itemize}
\item Sum performance: $f(\vect{g}) = \sum_k  g_k$;

\item Proportional fairness: $f(\vect{g}) = \prod_k g_k^{1/K_r}$;

\item Harmonic mean: $f(\vect{g}) = K_r ( \sum_k g_k^{-1} )^{-1}$;

\item Max-min fairness: $f(\vect{g}) = \min_k g_k$.
\end{itemize}
Weights can be included in these examples to compensate for heterogeneous channel conditions, delay constraints, etc.

Herein, we assume that the system performance function $f(g_1(\widetilde{\text{MSE}}_1),\ldots,g_{K_r}(\widetilde{\text{MSE}}_{K_r}))$ is Lipschitz continuous and \emph{strictly increasing}\footnote{A function $f: \mathbb{R}_+^n \rightarrow \mathbb{R}$ is \emph{strictly increasing} if for any $\vect{x},\vect{x'},\vect{x''} \in \mathbb{R}_+^n$ such that $\vect{x} \geq \vect{x'}$ and  $\vect{x} > \vect{x''}$, it follows $f(\vect{x})\geq f(\vect{x}')$ and $f(\vect{x})>f(\vect{x}'')$.}. This is satisfied by the aforementioned examples, and by all reasonable system performance measures. When combined with a performance region $\mathcal{R}$ that is compact and normal, we have the following important result.

\begin{lemma} \label{lemma_global_optimum_attained_on_boundary}
If $f(\cdot)$ is a strictly increasing function and $\mathcal{R}$ is a compact and normal set, the global optimum (if it exists) to
\begin{equation} \label{eq_global_problem}
\maximize{\vect{g} \in \mathcal{R} } \,\,\, f(\vect{g})
\end{equation}
is attained on $\partial^+ \mathcal{R}$. In addition, for any $\tilde{\vect{g}} \in \partial^+ \mathcal{R}$ there exists a strictly increasing $f(\cdot)$ with $\tilde{\vect{g}}$ as global optimum.
\end{lemma}
\begin{IEEEproof}
The first statement is proved in \cite[Proposition 7]{Tuy2000a}. The second statement is proved using the strictly increasing function $f(\vect{g}) = \min_k g_k / \tilde{g}_k$ with $\tilde{\vect{g}}=(\tilde{g}_1,\ldots,\tilde{g}_{K_r}) \in \partial^+ \mathcal{R}$. Obviously, $\max_{\vect{g} \in \mathcal{R} } f(\vect{g}) \geq f(\tilde{\vect{g}})=1$ and assume for the purpose of contradiction that it exists $\vect{g}^* \in \mathcal{R}$ that achieves strict inequality. This means that $\vect{g}^* > \tilde{\vect{g}}$ and thus $\tilde{\vect{g}}$ cannot be an upper boundary point since $\{\vect{y}' \in \mathbb{R}_+^{n}: \, \vect{y}'> \tilde{\vect{g}} \} \cap \mathcal{R} \neq \emptyset$ (see Definition \ref{def_upper_boundary}). This contradiction yields
$\max_{\vect{g} \in \mathcal{R} } f(\vect{g}) = f(\tilde{\vect{g}})$ and thus $\tilde{\vect{g}}$ is the (non-unique) global optimum.
\end{IEEEproof}

Based on this lemma, we only need to search the upper boundary of $\mathcal{R}$ to solve any system performance optimization problem. However, this is not as simple as it seems; \cite{Luo2008a} showed that sum performance maximization is NP-hard for any number of transmit antennas, while \cite{Liu2011a} showed NP-hardness for the harmonic mean and proportional fairness for $N_j>1$. A main characteristic of NP-hard problems is that there are no known algorithms that solve them in polynomial time, and it is widely believed that there exist no such algorithms.

From \cite{Luo2008a,Liu2011a} it is fair to say that system performance optimization is generally NP-hard. However, there is a useful problem that can be solved efficiently (i.e., in polynomial time), namely the max-min fairness optimization (defined above) \cite{Liu2011a}. It belongs to a larger category of problems, \emph{robust fairness-profile optimization}, that we analyze in Section \ref{section_opt_fairnessprofile} under channel uncertainty. It is also an essential subproblem of the BRB algorithm in Section \ref{section_monotonic_optimization} that solves the monotonic optimization problem for any $f(\cdot)$, although the NP-hardness makes the convergence unsuitable for real-time applications.

\section{Robust Fairness-Profile Optimization}
\label{section_opt_fairnessprofile}

In this section, we consider a particular $f(\cdot)$ for which \eqref{eq_global_problem} can be solved efficiently and which is used as subproblem of the general BRB algorithm in the next section. The considered robust system performance optimization problem is
\begin{equation} \label{eq_optproblem_performance-profile}
\begin{split}
\maximize{ \substack{ ( \vect{v}_1,\ldots,\vect{v}_{K_r} ) \in
\mathcal{V} \\ (r_1,\ldots,r_{K_r}) \in \mathbb{R}_+^{K_r} } }\,\, & \,\, \min_k \frac{g_k(\widetilde{\text{MSE}}_k) - a_k }{\alpha_k }, \\
\mathrm{subject}\,\,\mathrm{to}\,\,\,\,\,\,\, & \,\,\, g_k(\widetilde{\text{MSE}}_k) \geq a_k \quad \forall k.
\end{split}
\end{equation}
This problem can be seen as a generalization of classic robust max-min optimization (see e.g., \cite{Shenouda2009a}) where two fairness constraints have been added:
\begin{enumerate}
\item Each user has a lowest acceptable level $g_k(\widetilde{\text{MSE}}_k) \geq a_k$;
\item The total performance above this level is divided such that each user gets a predefined portion $\alpha_k \geq 0$.
\end{enumerate}
The first constraint is represented by $\vect{a}=[a_1,\ldots, a_{K_r}]^T \geq \vect{0}$. The second constraint is called a \emph{fairness-profile}\footnote{The terminology rate-profile has been used for similar problems in prior work \cite{Zhang2010b,Karipidis2010a,Mohseni2006a}, but herein we extend these works by having arbitrary performance measures, uncertain CSI, and general multicell scenarios.} and is symbolized by a vector $\boldsymbol{\alpha}=[\alpha_1,\ldots \alpha_{K_r}]^T$ that satisfies $\sum_{k=1}^{K_r} \alpha_k=1$ (without loss of generality).

\begin{figure}[t]
\begin{center}
\includegraphics[width=\columnwidth]{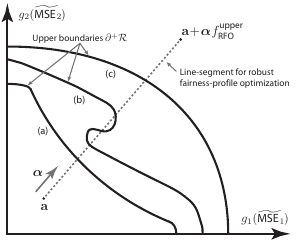}
\end{center} \vskip -5mm
\caption{Examples of robust performance regions with different shapes. (a) Region is normal but non-convex. (b) Region is neither normal nor convex. (c) Region is both normal and convex. Simple bisection along a fairness-profile is not guaranteed to find the upper boundary of non-normal regions.}\label{figure_region_examples} \vskip -5mm
\end{figure}

We call \eqref{eq_optproblem_performance-profile} a \emph{robust fairness-profile optimization} (RFO) and observe that this problem has a simple geometrical interpretation; we start in
$\vect{a} \in \mathcal{R}$ and follow a ray in the direction of $\boldsymbol{\alpha}$ until a point on the upper boundary $\partial^+ \mathcal{R}$ is found.\footnote{This geometrical approach finds an optimal solution to \eqref{eq_optproblem_performance-profile} where $(g_k(\widetilde{\text{MSE}}_k) \!-\! a_k)/\alpha_k$ is the same for all $\text{MS}_k$. In certain special cases (e.g., when the upper boundary is flat in some dimension), there also exist solutions where a few users get higher performance than this worst-user level. This discussed in \cite{Maddah2009a}.}
In general search regions, the ray might leave the region and come back again which makes the search very complicated.
Fortunately, $\mathcal{R}$ is a compact and normal set and thus the ray intersects the upper boundary in a unique point.
This is illustrated in Fig.~\ref{figure_region_examples}, where (a) and (c) are normal sets while (b) is non-normal and thus some rays from within the set cross the upper boundary multiple times.

If we can find an upper bound $f^{\text{upper}}_{\text{RFO}}$ on the optimal value of \eqref{eq_optproblem_performance-profile}, we know geometrically that the optimum lies on the line-segment between $\vect{a}$ and $\vect{a}+ \boldsymbol{\alpha} f^{\text{upper}}_{\text{RFO}}$; see the illustration in Fig.~\ref{figure_region_examples}. Hoping to simplify the RFO problem, we can thus rewrite \eqref{eq_optproblem_performance-profile} as a bisection over this line-segment.

\begin{lemma} \label{lemma_fairness_profile_bisection}
For compact uncertainty sets $\mathcal{U}_1,\ldots,\mathcal{U}_{K_r}$, fixed $\vect{a},\boldsymbol{\alpha}$ and a given upper bound $f^{\text{upper}}_{\text{RFO}}$ on the optimum of \eqref{eq_optproblem_performance-profile}, the problem can be solved by bisection over the range $\mathcal{F}=[0,f^{\text{upper}}_{\text{RFO}}]$. For a given $f^{\text{candidate}}_{\text{RFO}} \in \mathcal{F}$, the feasibility problem
\begin{equation} \label{eq_subproblem_performance-profile}
\begin{split}
\mathrm{find}\,\, & \,\, \vect{v}_1,\ldots,\vect{v}_{K_r},r_1\geq 0,\ldots,r_{K_r}\geq 0 \\
\mathrm{subject}\,\,\mathrm{to}\,\, & \,\, \widetilde{\text{MSE}}_k \leq \gamma_k \,\, \forall k, \\ & \, \,
 \sum_{k} \vect{v}_k^H \vect{Q}_l \vect{v}_k \leq q_l \quad \forall l
\end{split}
\end{equation}
is solved for $\gamma_k=g_k^{-1} (a_k \!+\! \alpha_k f^{\text{candidate}}_{\text{RFO}})$.
If the problem is feasible, all $\tilde{f} \in \mathcal{F}$ with $\tilde{f}< f^{\text{candidate}}_{\text{RFO}}$ are removed. Otherwise, all $\tilde{f} \in \mathcal{F}$ with $\tilde{f} \geq f^{\text{candidate}}_{\text{RFO}}$ are removed. The initial feasibility of \eqref{eq_optproblem_performance-profile} is checked by solving \eqref{eq_subproblem_performance-profile} for $f^{\text{candidate}}_{\text{RFO}}=0$.
\end{lemma}
\begin{IEEEproof}
From Lemma \ref{lemma_performance_region_normal}, $\mathcal{R}$ is a compact and normal set. For such sets, a ray from a point within the region (in a positive direction) meets $\partial^+ \mathcal{R}$ in a unique point (see \cite[Proposition 6]{Tuy2000a}). This point is the optimum to \eqref{eq_optproblem_performance-profile}, since the optimum must be on $\partial^+ \mathcal{R}$ (as proved in Lemma \ref{lemma_global_optimum_attained_on_boundary}). As the ray only meets the upper boundary once, it can be divided into two parts: one part is inside of $\mathcal{R}$ and one part is outside. The intersection can be found (to any accuracy) by a line search (e.g., bisection) that iteratively checks if a point $\vect{a} \!+\! \boldsymbol{\alpha} f^{\text{candidate}}_{\text{RFO}}$ is inside $\mathcal{R}$ by solving \eqref{eq_subproblem_performance-profile}.
\end{IEEEproof}

Obviously, the RFO problem in \eqref{eq_optproblem_performance-profile} is infeasible if $\vect{a}$ is outside of $\mathcal{R}$, which can be checked as described in the lemma. A successful bisection also requires an initial selection of $f^{\text{upper}}_{\text{RFO}}$ in Lemma \ref{lemma_fairness_profile_bisection} such that $\vect{a} \!+\! \boldsymbol{\alpha} f^{\text{upper}}_{\text{RFO}}$ is outside $\mathcal{R}$. If not given in advance, $f^{\text{upper}}_{\text{RFO}}$ can be achieved in different ways:
\begin{itemize}
\item $f^{\text{upper}}_{\text{RFO}} = K_r \, \mathrm{sup}_k (g_k(0)-a_k)$. Cannot be used if $\mathrm{sup}_k \, g_k(0)=\infty$.

\item $f^{\text{upper}}_{\text{RFO}} = \sum_k g_k(\sigma_k^2/( \kappa_k \| \vect{D}_k^H \widehat{\vect{h}}_k\|_2^2+\sigma_k^2))-a_k$, where $\kappa_k$ is a bound on the transmit power and is calculated as the smallest positive eigenvalue of $\frac{\vect{D}_k^H \vect{Q}_l \vect{D}_k}{q_l \tr(\vect{D}_k)}$ among all $l$.

\item $f^{\text{upper}}_{\text{RFO}} = \sum_k g_k(\widetilde{\textrm{MSE}}_{\text{su},k})-a_k$, where $\widetilde{\textrm{MSE}}_{\text{su},k}$ is the optimal robust MSE if $\text{MS}_k$ is the only active user.
\end{itemize}
The first one is the simplest and ignores the power constraints, while the second one ignores co-user interference and uncertainty and assumes that the highest power available in some spatial direction can be used in any direction. The third one takes the MSEs achieved in a single-user system and requires that these problems are solved (which is simple under some power constraints), but achieves the tightest value on $f^{\text{upper}}_{\text{RFO}}$.

\subsection{Convexity of Feasibility Subproblems}
\label{subsection_feasibility_problems}

Solving the RFO problem using bisection, as suggested in Lemma \ref{lemma_fairness_profile_bisection}, is appealing as the range is halved in each iteration; thus, the number of iterations scales only logarithmical with the desired accuracy $\delta$ of the solution, also known as linear convergence. In other words, the computational complexity is typically not limited by the number of iterations but by the complexity of the feasibility problem \eqref{eq_subproblem_performance-profile} solved in each iteration. Next, we will see that \eqref{eq_subproblem_performance-profile} can be solved efficiently.

If the transmitters have perfect CSI (i.e., $\mathcal{U}_k=\{\widehat{\vect{h}}_k\}$), the feasibility problem in \eqref{eq_subproblem_performance-profile} is convex \cite{Shenouda2009a} and can be efficiently solved (e.g., using general-purpose implementations of interior-point methods \cite{SDPT3}); see Appendix \ref{appendix_perfect_CSI} for further details. Under worst-case robustness to CSI uncertainty, the feasibility problem in \eqref{eq_subproblem_performance-profile} seems difficult to solve since there are infinitely many MSE constraints (one for each $\vect{h}_k \in \mathcal{U}_k$). Fortunately, the following theorem provides a reformulation into finitely many convex constraints, based upon well-known results from robust optimization \cite{Bental2009a}.

\begin{theorem} \label{theorem_convex_feasibility_problem}
For the compact uncertainty sets in \eqref{eq_def_uncertainty_sets}, the feasibility problem in \eqref{eq_subproblem_performance-profile} is equivalent to the convex problem
\begin{equation} \label{eq_expressed_as_sdp}
\begin{split}
\mathrm{find}\,\, & \,\, \vect{v}_1,\ldots,\vect{v}_{K_r},\tilde{r}_1\geq 0,\ldots,\tilde{r}_{K_r}\geq 0,\\ & \,\, \lambda_1\geq 0,\ldots,\lambda_{K_r}\geq 0 \\
\mathrm{subject}\,\,\mathrm{to}\,\, & \,\,
\vect{A}_k \succeq \vect{0}_{N+K_r+2} \,\, \forall k, \\ & \, \,
 \sum_{k} \vect{v}_k^H \vect{Q}_l \vect{v}_k \leq q_l \quad \forall l
\end{split}
\end{equation}
where $\widetilde{\vect{V}}=[\vect{D}_1 \vect{v}_1 \, \ldots \, \vect{D}_{K_r} \vect{v}_{K_r}]$ and
\begin{equation} \label{eq_Ak-matrix}
\begin{split}
&
\vect{A}_k= \\ &
\left[\begin{IEEEeqnarraybox*}[][c]{cccc} \!
\sqrt{\gamma_k} \tilde{r}_k \!-\! \lambda_k & \widehat{\vect{h}}_k^H \vect{C}_k \widetilde{\vect{V}} \!-\!\tilde{r}_k \vect{e}_k^T & \sigma_k & \vect{0} \\
\widetilde{\vect{V}}^H \vect{C}_k^H \widehat{\vect{h}}_k \!-\!\tilde{r}_k \vect{e}_k & \sqrt{\gamma_k} \tilde{r}_k \vect{I}_{K_r} & \vect{0} & -\widetilde{\vect{V}}^H \vect{C}_k^H \vect{B}_k \\
\sigma_k & \vect{0} & \sqrt{\gamma_k} \tilde{r}_k & \vect{0} \\
\vect{0} & -\vect{B}_k^H \vect{C}_k  \widetilde{\vect{V}} & \vect{0} & \lambda_k \vect{I}_{N}%
\end{IEEEeqnarraybox*} \! \right]\!.
\end{split}
\end{equation}
\end{theorem}
\begin{IEEEproof}
The proof is given in Appendix \ref{appendix_robust_reformulation}.
\end{IEEEproof}

This theorem only has one (linear) semi-definite constraint per user and has replaced the uncertainty set $\mathcal{U}_k$ by a variable $\lambda_k$ that indirectly represents the worst channel;
if we can find $\lambda_k\geq 0$ that satisfies the constraint, then the original MSE constraints are satisfied for all $\vect{h}_k \in \mathcal{U}_k$.

Single-cell counterparts to Theorem \ref{theorem_convex_feasibility_problem} have recently been derived in \cite{Zheng2008b,Vucic2009a,Shenouda2009a}, while the multicell generalization is novel. Special cases of the fairness-profile optimization problem have also been considered before; if $g_k(\text{MSE}_k)=\text{MSE}_k^{-1}-1$ and $\vect{a}=\vect{0}$, the problem is equivalent to the minimization of the (weighted) worst MSE among all users \cite{Schubert2004a,Wiesel2006a,Vucic2009a,Shenouda2009a}. This special case can be posed as a generalized eigenvalue problem \cite{Boyd1993a,Wiesel2006a,Shenouda2009a}, which can improve the computational complexity. For general user utility functions $g_k(\cdot)$, such simplification is not possible.

The bisection algorithm for \eqref{eq_optproblem_performance-profile} is summarized in Table \ref{algorithm_robust_performanceprofile}.

\begin{table} [ht!]
\vskip -1mm \caption{Algorithm 1: Robust Fairness-Profile Optimization} \vskip -5mm
\label{algorithm_robust_performanceprofile}
\begin{lined}{8.4 cm} \vskip -1mm
\renewcommand{\labelenumi}{\theenumi:}
\begin{enumerate}

\item \textbf{input} starting-point $\vect{a}$ and fairness-profile $\boldsymbol{\alpha}$

\item \textbf{input} accuracy $\delta$, $f^{\text{lower}}_{\text{RFO}}=0$, and $f^{\text{upper}}_{\text{RFO}}$ (see suggestions)

\item \textbf{while} $f^{\text{upper}}_{\text{RFO}}-f^{\text{lower}}_{\text{RFO}}>\delta$

\item $\,\,\,$ set $f^{\text{candidate}}_{\text{RFO}} = (f^{\text{lower}}_{\text{RFO}}+f^{\text{upper}}_{\text{RFO}})/2$

\item $\,\,\,$ set $\gamma_k=g_k^{-1} (a_k \!+\! \alpha_k f^{\text{candidate}}_{\text{RFO}}) \quad \forall k$

\item $\,\,\,\,\,$\textbf{if} problem \eqref{eq_subproblem_performance-profile} is
feasible for these $\gamma_k$: set $f^{\text{lower}}_{\text{RFO}} = f^{\text{candidate}}_{\text{RFO}}$

\item $\,\,\,\,\,$\textbf{else}: set $f^{\text{upper}}_{\text{RFO}} =
f^{\text{candidate}}_{\text{RFO}} \,\,\,$  \textbf{end}

\item \textbf{end}

\item \textbf{return} $[f^{\text{lower}}_{\text{RFO}},f^{\text{upper}}_{\text{RFO}}]$ and last feasible solution to
\eqref{eq_subproblem_performance-profile} from step 6
\end{enumerate}
\vskip 2mm
\end{lined}\vskip-4mm
\end{table}

\subsection{Complexity of Feasibility Subproblems}

The previous section showed that the feasibility problem \eqref{eq_subproblem_performance-profile} can be expressed as a convex problem under both perfect CSI and worst-case robustness.
Thus, it can be solved with a computational complexity that is polynomial in the number of antennas $N$, users $K_r$, and power constraints $L$ \cite[Chapter 6]{Bental2001a}.
The exact complexity depends both on current systems conditions and the choice of solver algorithm (e.g., interior-point methods \cite{SDPT3}), but we illustrate the complexity in Section \ref{section_numerical_examples}.

Having CSI uncertainty will naturally increase the computational complexity, because Theorem \ref{theorem_convex_feasibility_problem} handles the infinitely many MSE constraints by introducing
extra variables $\lambda_k$ and because the resulting MSE constraints have larger dimension than under perfect CSI. In addition, the problem size under perfect CSI (and thereby the complexity) can be reduced by plugging in the optimal equalizing coefficients and rewriting the corresponding MSE expressions as second-order cone constraints \cite{Wiesel2006a,Yu2007a}; see Appendix \ref{appendix_perfect_CSI}. Fixed point algorithms can provide fast solutions under perfect CSI with total power constraints \cite{Schubert2005a}, but need to be combined with outer optimization procedures under general power constraints \cite{Yu2007a}. CSI uncertainty will, on the other hand, reduce the size of the performance region $\mathcal{R}$, thus fewer feasibility problems need to be solved to attain a given accuracy $\delta$.

Finally, note that Lemma \ref{lemma_fairness_profile_bisection} solves $\lceil \log_2(f^{\text{upper}}_{\text{RFO}}/\delta)\rceil$ feasibility problems to achieve the prescribed accuracy $\delta$ on the solution to the RFO problem. This number is bounded by a constant, therefore the RFO problem also have polynomial complexity in the number of antennas $N$, users $K_r$, and power constraints $L$. This complexity is quite affordable, making the RFO problem a reasonable candidate for resource allocation in practical systems. To put it differently, the system designer basically has the choice between solving a RFO problem optimally or solving some other NP-hard resource allocation problem \eqref{eq_global_problem} suboptimally.

\subsection{Extensions to the System Model}
\label{subsection_extensions}

The RFO approach is easily extended to any robustness scenario where the following properties are satisfied:
\begin{itemize}
\item The performance region is compact and normal;
\item The feasibility problem in \eqref{eq_subproblem_performance-profile} can be solved efficiently.
\end{itemize}
The first property is usually satisfied; observe that Lemma \ref{lemma_performance_region_normal} proved it for any compact uncertainty sets (not only ellipsoidal sets). However, it is often difficult to solve the feasibility problem efficiently under general uncertainty sets. Tractability can be achieved through conservative approximations that give lower bounds on performance, see \cite{Shenouda2009a,Vucic2009a} for examples with rectangular sets and intersections between ellipsoidal sets. If the worst-case robustness is replaced with probabilistic robustness constraints,
conservative approximations of each user's performance are required to achieve tractable problem formulations \cite{Chalise2007a,Shenouda2008a,Wang2011a}. The probabilistic approach enables user performance functions based on the outage probability and outage data rate.
Other possible extensions to the system model is Tomlinson-Harashima precoding (see \cite{Shenouda2009a}), soft-shaping constraints (see \cite{Scutari2008a}), and multi-carrier systems (see \cite{Bjornson2011a}).

\section{Robust Monotonic Optimization}
\label{section_monotonic_optimization}

In this section, we aim at solving the robust monotonic optimization problem in \eqref{eq_global_problem} for any system performance function. Recall from Lemma \ref{lemma_global_optimum_attained_on_boundary} that the optimum lies on the upper boundary. Hence, we can in principle look for an approximate solution in a large set of boundary points of $\mathcal{R}$ achieved by solving the RFO problem in Section \ref{section_opt_fairnessprofile} for a very fine grid of fairness-profiles $\boldsymbol{\alpha}$. However, this naive approach has huge computational complexity, which calls for a more systematic approach that concentrates on the boundary in \emph{good} directions.

Next, we propose a branch-reduce-and-bound (BRB) algorithm for solving \eqref{eq_global_problem} systematically and with global convergence.
It can be seen as an adaptation of the generic BRB algorithm in \cite{Tuy2005a} to general multicell transmission.

The algorithm maintains a set $\mathcal{N}$ with non-overlapping hyperrectangles that surely covers the parts of the robust performance region $\mathcal{R}$ where the optimal solutions lie (the solution might be non-unique). Iteratively, we split certain hyperrectangles and try to improve a lower bound $f_{\min}$ and an upper bound $f_{\max}$ on the optimal value of \eqref{eq_global_problem}. To aid this process, a local upper bound $\beta(\mathcal{M})$ is also stored for each $\mathcal{M} \in \mathcal{N}$. The algorithm proceeds until $f_{\max}-f_{\min}<\varepsilon$, for a predefined solution accuracy $\varepsilon$.

In what follows, hyperrectangles are called $\emph{boxes}$:
\begin{definition}
For given $\vect{a},\vect{b} \in \mathbb{R}^{K_r}_+$ with $\vect{a} \leq \vect{b}$, the set of all $\vect{x}$ such that $\vect{a} \leq \vect{x} \leq\vect{b}$ is called a \emph{box} and is denoted $[\vect{a},\vect{b}]$.
\end{definition}

Initially, $\mathcal{N}=\{\mathcal{M}_0\}$ for a box $\mathcal{M}_0=[\vect{0},\vect{b}_0] \subset \mathbb{R}_+^{K_r}$ where $\vect{b}_0$ is based on some suitable upper bound that guarantees $\mathcal{R} \subseteq \mathcal{M}_0$ (see suggestions in Section \ref{section_opt_fairnessprofile}). The initial lower and upper bounds can be taken as $f_{\min}=f(\vect{0})=0$ and $f_{\max}=f(\vect{b}_0)$, but some low-complexity resource allocation strategy can be used to obtain a better lower bound.

\begin{figure}[t!]
\begin{center}
\includegraphics[width=\columnwidth]{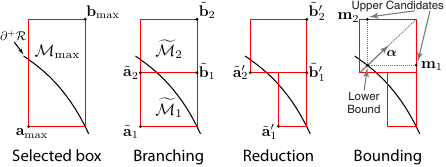}
\end{center} \vskip -5mm
\caption{An iteration of the BRB algorithm: A box is selected and branched into two new boxes. These are reduced based on the current bounds on the optimal value. Finally, line search between the lower and upper corners of the outmost box is applied to improve the bounds.}\label{figure_brb_algoritm} \vskip -3mm
\end{figure}

Each iteration of the BRB algorithm consists of three steps.
\begin{enumerate}
\item Branching: Divide a box from $\mathcal{N}$ into two new boxes.
\item Reduction: Remove parts of these new boxes that cannot contain optimal solutions.
\item Bounding: Search for a feasible solution in one of the new boxes and use it to improve $f_{\min}$, $f_{\max}$, and $\beta(\cdot)$.
\end{enumerate}

These steps are illustrated in Fig.~\ref{figure_brb_algoritm} and the details are explained in the next subsections. The final algorithm is given in Section \ref{subsection_brb_algorithm}. For notational convenience, the $k$th element of any vector $\vect{x}$ is denoted $x_k$.

\subsection{Branching}

Each iteration begins with selecting a box $\mathcal{M}_{\max}=[\vect{a}_{\max},\vect{b}_{\max}]$ that contains the current upper bound $f_{\max}$:
\begin{equation}
\mathcal{M}_{\max} = \argmax{\mathcal{M} \in \mathcal{N}} \,\, \beta (\mathcal{M}).
\end{equation}
The intention is to improve the upper bound by partitioning $\mathcal{M}_{\max}$ into two new boxes $\widetilde{\mathcal{M}}_1,\widetilde{\mathcal{M}}_2$.
New boxes of equal size are achieved by bisecting $\mathcal{M}_{\max}$ along its longest side (see Fig.~\ref{figure_brb_algoritm}). The index of this side is $\mathrm{dim} = \mathrm{argmax}_k (b_{\max,k}-a_{\max,k})$ and the new boxes are
\begin{equation} \label{eq_new_boxes}
\begin{split}
\widetilde{\mathcal{M}}_1 &= [\vect{a}_{\max},\vect{b}_{\max}-s \vect{e}_{\mathrm{dim}} ], \\
\widetilde{\mathcal{M}}_2 &= [\vect{a}_{\max}+s \vect{e}_{\mathrm{dim}},\vect{b}_{\max} ],
\end{split}
\end{equation}
where $s=(b_{\max,\mathrm{dim}}-a_{\max,\mathrm{dim}})/2$ and $\vect{e}_{k}$ is the $k$th column of $\vect{I}_{K_r}$. The (local) upper bounds over these new boxes are also based on $\mathcal{M}_{\max}$:
\begin{equation} \label{eq_new_boxes_upper_bounds}
\begin{split}
\beta(\widetilde{\mathcal{M}}_1) &= \min(\beta(\mathcal{M}_{\max}),f(\vect{b}_{\max}-s \vect{e}_{\mathrm{dim}})), \\
\beta(\widetilde{\mathcal{M}}_2) &= \beta(\mathcal{M}_{\max}).
\end{split}
\end{equation}
Finally, the set $\mathcal{M}_{\max}$ is replaced with $\widetilde{\mathcal{M}}_1$ and $\widetilde{\mathcal{M}}_2$ in $\mathcal{N}$.

\subsection{Reduction}
In this step, the new boxes $\widetilde{\mathcal{M}}_l=[\tilde{\vect{a}}_l,\tilde{\vect{b}}_l]$, for $l=1,2$, are reduced by cutting off parts that cannot achieve function values between the lower bound $f_{\min}$ and (local) upper bound $\beta(\widetilde{\mathcal{M}}_l)$. If $\beta(\widetilde{\mathcal{M}}_l)<f_{\min}$, the whole box is removed from $\mathcal{N}$. Otherwise, it is replaced by a (potentially) smaller box $[\tilde{\vect{a}}'_l,\tilde{\vect{b}}'_l]$ based on the following lemma.

\begin{lemma} \label{lemma_reduction_algorithm}
If $f_{\min} \leq \beta(\widetilde{\mathcal{M}}_l)$, all points $\vect{g} \in [\tilde{\vect{a}}_l,\tilde{\vect{b}}_l]$ satisfying $f_{\min} \leq f(\vect{g}) \leq  \beta(\widetilde{\mathcal{M}}_l)$ are also contained in $[\tilde{\vect{a}}'_l,\tilde{\vect{b}}'_l] \subseteq [\tilde{\vect{a}}_l,\tilde{\vect{b}}_l]$, where
\begin{align}
\tilde{\vect{a}}'_l &= \tilde{\vect{b}}_l - \sum_{k=1}^{K_r} \nu_k  ( \tilde{b}_{l,k}- \tilde{a}_{l,k}) \vect{e}_k \label{eq_reduce_from_below} \\
\tilde{\vect{b}}'_l &= \tilde{\vect{a}}'_l + \sum_{k=1}^{K_r} \mu_k  ( \tilde{b}_{l,k}- \tilde{a}'_{l,k}) \vect{e}_k \label{eq_reduce_from_above}
\end{align}
with $\nu_k$ and $\mu_k$ (for $k=1,\ldots,K_r$) calculated as
\begin{equation} \label{eq_reduction_variables}
\begin{split}
\nu_k &\!=\! \max \left\{ \nu: 0 \!\leq\! \nu \!\leq\! 1, f( \tilde{\vect{b}}_l - \nu (\tilde{b}_{l,k}-\tilde{a}_{l,k}) \vect{e}_k ) \geq f_{\min} \right\} \\
\mu_k &\!=\! \max \left\{ \mu: 0 \!\leq\! \mu \!\leq\! 1, f( \tilde{\vect{a}}'_l + \mu (\tilde{b}_{l,k}-\tilde{a}'_{l,k}) \vect{e}_k ) \leq \beta(\widetilde{\mathcal{M}}_l) \right\} \!.
\end{split}
\end{equation}
\end{lemma}
\begin{IEEEproof}
The proof is given in Appendix \ref{appendix_reduction_algorithm}.
\end{IEEEproof}

The reduction procedure in Lemma \ref{lemma_reduction_algorithm} is illustrated in Fig.~\ref{figure_brb_algoritm} and observe that it needs to be implemented sequentially; first, the lower point $\tilde{\vect{a}}_l$ is updated using \eqref{eq_reduce_from_below} and then it is used to update the upper point $\tilde{\vect{b}}_l$ using \eqref{eq_reduce_from_above}.

Each reduction requires calculation of the parameters $\nu_k,\mu_k$ in \eqref{eq_reduction_variables}, generally solved by standard line search procedures. However, closed form expressions can be attained in many cases. For example, weighted sum performance with $f(\vect{g}) = \sum_k w_k g_k$ (with weights $w_k>0$) gives
\begin{equation}
\begin{split}
\nu_k &= \min\left( \frac{\sum_{\bar{k}=1}^{K_r} w_{\bar{k}} \tilde{b}_{l,k} - f_{\min}}{w_k ( \tilde{b}_{l,k}- \tilde{a}_{l,k}) }, 1 \right), \\
\mu_k &= \min\left( \frac{\beta(\widetilde{\mathcal{M}}_l) - \sum_{\bar{k}=1}^{K_r} w_{\bar{k}} \tilde{a}'_{l,k}  }{w_k ( \tilde{b}_{l,k}- \tilde{a}'_{l,k}) }, 1 \right)
\end{split}
\end{equation}
where the min-operator makes sure that $\nu_k,\mu_k\leq 1$.

\subsection{Bounding}

Each iteration ends with a bounding step where we search for feasible solutions in $\widetilde{\mathcal{M}}_2=[\tilde{\vect{a}}'_2,\tilde{\vect{b}}'_2]$, which is the new box with the largest (local) upper bound (i.e., $\beta(\widetilde{\mathcal{M}}_2) \geq \beta(\widetilde{\mathcal{M}}_1)$). These solutions are used to improve $f_{\min}$, $f_{\max}$, and $\beta(\widetilde{\mathcal{M}}_2)$.

First, the feasibility of the lower corner $\tilde{\vect{a}}'_2$ is checked by solving \eqref{eq_subproblem_performance-profile} with $\gamma_k=g_k^{-1} ( \tilde{a}'_{2,k} ) \,\, \forall k$. If this problem is infeasible, then $\widetilde{\mathcal{M}}_2$ contains no feasible solutions and is removed from $\mathcal{N}$. If the problem is feasible, the following lemma is used to find lower and upper bounds on the feasible performance in $\widetilde{\mathcal{M}}_2$ by solving a single robust fairness-profile optimization problem (see Section \ref{section_opt_fairnessprofile}):

\begin{lemma} \label{lemma_bounding_algorithm}
Consider a box $\mathcal{M}=[\vect{a},\vect{b}] \subset \mathbb{R}^{K_r}_+$ such that $\mathcal{M} \cap \mathcal{R} \neq \emptyset$. If $\mathcal{R}$ is compact and normal, the highest feasible performance in $\mathcal{M}$ can be bounded as $[\bar{f}_{\min},\bar{f}_{\max}]$ for
\begin{equation}
\begin{split}
\bar{f}_{\min} &= f(\vect{a}+ \boldsymbol{\alpha} f^{\min}_{\text{RFO}}) \\
\bar{f}_{\max} &= \max_{k} f( \vect{b} - (b_{k} - n_k) \vect{e}_k )
\end{split}
\end{equation}
where $\vect{n}=[n_1,\ldots,n_{K_r}]^T=\vect{a}+\boldsymbol{\alpha} f^{\max}_{\text{RFO}}$, $\boldsymbol{\alpha}=(\vect{b}-\vect{a})/\| \vect{b}-\vect{a}\|_1$, and $\vect{e}_k$ denotes the $k$th column of $\vect{I}_{K_r}$. The variables $f^{\min}_{\text{RFO}},f^{\max}_{\text{RFO}}$ are the interval endpoints achieved by Algorithm 1 with starting-point $\vect{a}$, fairness-profile $\boldsymbol{\alpha}$, $f^{\text{upper}}_{\text{RFO}}=\| \vect{b} - \vect{a}\|_1$, and some given line-search accuracy $\delta$.
\end{lemma}
\begin{IEEEproof}
The proof is given in Appendix \ref{appendix_bounding_algorithm}.
\end{IEEEproof}

The lemma is illustrated in Fig.~\ref{figure_brb_algoritm}, where a line-search is performed between the lower and upper corner of the box. The best feasible point on this line provides a local lower bound $\bar{f}_{\min}$ on the feasible performance. Since the region is normal, the outer points $\vect{m}_k=\tilde{\vect{b}}'_2 - (\tilde{b}'_{2,k} - n_k) \vect{e}_k$ are candidates for giving a new upper bound on the feasible performance in the box. Observe that if the size of the box $\|\tilde{\vect{b}}'_2-\tilde{\vect{a}}'_2\|_1$ is smaller than the accuracy $\delta$, then no line-search is performed. This will not affect the convergence, as proved in the next subsection.

The local lower bound $\bar{f}_{\min}$ from Lemma \ref{lemma_bounding_algorithm} replaces the global lower bound $f_{\min}$ if $\bar{f}_{\min} \geq f_{\min}$.
Similarly, we set $\beta(\widetilde{\mathcal{M}}_2) = \bar{f}_{\max}$ if $\bar{f}_{\max} < \beta(\widetilde{\mathcal{M}}_2)$. Finally, we update $f_{\max}$ with the largest upper bound $\max_{\mathcal{M} \in \mathcal{N}} \beta(\mathcal{M})$ among the remaining boxes. The stopping criterion $f_{\max}-f_{\min}<\varepsilon$ is checked before a new iteration is started.

\subsection{Final Algorithm}
\label{subsection_brb_algorithm}

The BRB algorithm that solves the general robust monotonic optimization problem in \eqref{eq_global_problem} is summarized in Table \ref{algorithm_brb}.

\begin{table}[thb]
\vskip -1mm \caption{Algorithm 2: Branch-Reduce-and-Bound} \vskip -5mm
\label{algorithm_brb}
\begin{lined}{8.2 cm} \vskip -1mm
\renewcommand{\labelenumi}{\theenumi:}
\begin{enumerate}

\item \textbf{input} $\mathcal{M}_0\!=\![\vect{0},g_{\text{max}}\vect{1}_{K_r}]$, accuracy $\varepsilon$, line-search accuracy $\delta$

\item set $f_{\min},f_{\max}$ based on the available prior knowledge

\item set $\mathcal{N}=\{ \mathcal{M}_0 \}$ and $\beta(\mathcal{M}_0)=f_{\max}$

\item \textbf{while} $f_{\max}-f_{\min}>\varepsilon$

\item $\,\,\,$ set $\mathcal{M}_{\max} = \textrm{argmax}_{\mathcal{M} \in \mathcal{N}} \,\, \beta (\mathcal{M})$

\item $\,\,\,$ \textbf{for} $l=1,2$:

\item $\,\,\,\,\,\,$ create new box $\widetilde{\mathcal{M}}_l$ using \eqref{eq_new_boxes} and set $\beta(\widetilde{\mathcal{M}}_l)$ using \eqref{eq_new_boxes_upper_bounds}

\item $\,\,\,\,\,\,$ \textbf{if} $f_{\min} \leq \beta(\widetilde{\mathcal{M}}_l)$: reduce $\widetilde{\mathcal{M}}_l$ using Lemma \ref{lemma_reduction_algorithm}

\item $\,\,\,\,\,\,\,$\textbf{else}: set $\widetilde{\mathcal{M}}_l= \emptyset \,\,\,$ \textbf{end}

\item $\,\,\,$ \textbf{end}

\item $\,\,\,$ Check feasibility of $\tilde{\vect{a}}'_2$ by solving \eqref{eq_subproblem_performance-profile} for $\gamma_k=g_k^{-1} ( \tilde{a}'_{2,k} )$

\item $\,\,\,$ \textbf{if} feasible:

\item $\,\,\,\,\,\,$ Calculate bounds $\bar{f}_{\min},\bar{f}_{\max}$ in $\widetilde{\mathcal{M}}_2$ using Lemma \ref{lemma_bounding_algorithm}

\item $\,\,\,\,\,\,$ set $f_{\min} = \max (f_{\min},\bar{f}_{\min})$

\item $\,\,\,\,\,\,$ set $\beta(\widetilde{\mathcal{M}}_2)=\min(\beta(\widetilde{\mathcal{M}}_2),\bar{f}_{\max})$

\item $\,\,\,\,\,$\textbf{else}: set $\widetilde{\mathcal{M}}_2= \emptyset \,\,\,$ \textbf{end}

\item $\,\,\,$ set  $\mathcal{N} = (\mathcal{N} \setminus \mathcal{M}_{\max}) \cup \{\widetilde{\mathcal{M}}_1,\widetilde{\mathcal{M}}_2 \}$

\item $\,\,\,$ set $f_{\max} = \max_{\mathcal{M} \in \mathcal{N}} \beta(\mathcal{M})$

\item \textbf{end}

\item \textbf{return} $[f_{\min},f_{\max}]$ and a feasible solution that achieved $f_{\min}$
\end{enumerate}
\vskip 2mm
\end{lined}\vskip-4mm
\end{table}

 The BRB algorithm converges to the global optimum $f_{\text{opt}}$ in the sense that an $\varepsilon$-approximate interval $f_{\text{opt}}\in [f_{\min},f_{\max}]$, with $f_{\max}-f_{\min} \leq \varepsilon$, is achieved in finitely many iterations for any $\varepsilon>0$. The line-search accuracy $\delta$ is used in the bounding step to improve convergence speed, but there are no constraints on it to achieve convergence.

\begin{theorem}
For any given accuracy $\varepsilon>0$, the BRB algorithm finds an interval $[f_{\min},f_{\max}]$ for the optimal value of \eqref{eq_global_problem} that satisfies $f_{\max}-f_{\min} \leq \varepsilon$, in a finite number of iterations. The line-search accuracy $\delta>0$ can be selected arbitrarily.
\end{theorem}
\begin{IEEEproof}
The convergence of the algorithm in Table \ref{algorithm_brb} can be studied as a standard branch-and-bound algorithm, treating the reduction step (which does not remove the solution) as part of the bounding step. In the appendix of \cite{Balakrishnan1991a}, two sufficient conditions are given for achieving an $\varepsilon$-approximate solution in a finite number of iterations: 1) The bounding step truly calculates lower and upper bounds on the optimal value; 2) The difference $f_{\max}-f_{\min}$ converges (uniformly) to zero. The first condition was proved in Lemma \ref{lemma_bounding_algorithm}, while the second condition follows from the exhaustiveness of bisection and the Lipschitz continuity of $f$ (i.e., $\|\vect{b}-\vect{a}\| \leq \mathrm{constant}_1$ means that $f(\vect{b})-f(\vect{a}) \leq \mathrm{constant}_2$). Finally, observe that bounding the performance in a box using only the lower and upper corners satisfies these conditions; thus, any $\delta>0$ can be used.
\end{IEEEproof}

Although the algorithm converges, the worst-case convergence speed is exponential in the number of users $K_r$ because the problem is NP-hard.\footnote{Observe that the RFO problem avoids the exponential complexity by searching along one-dimensional curves in the $K_r$-dimensional user space, while the BRB algorithm considers all dimensions.} On the other hand, $N$ and $L$ are not affecting the convergence scaling of the BRB algorithm. The main computational complexity lies in the feasibility problem \eqref{eq_subproblem_performance-profile}, which is solved individually in Step 11 and as part of Lemma \ref{lemma_bounding_algorithm} (a RFO problem) in Step 13. Under channel uncertainty, a convex formulation was given in Theorem \ref{theorem_convex_feasibility_problem}. The dimension of the feasibility problem can be reduced under perfect CSI and other special conditions (see Appendix \ref{appendix_perfect_CSI} and Section \ref{section_opt_fairnessprofile}), but as the BRB algorithm solves a long sequence of convex subproblems the total complexity makes it unsuitable for real-time applications. However, we show numerically in Section \ref{section_numerical_examples} that the proposed algorithm has far better convergence behavior than the outer polyblock approximation in \cite{Brehmer2009a,Jorswieck2010a,Brehmer2010a,Utschick2012a}.

\begin{remark}
The BRB algorithm in Table \ref{algorithm_brb} is formulated to be applicable to any robust monotonic optimization problem. If only a certain type of user and system performance functions is of interest, this knowledge can be used to improve convergence. In particular, the bounding step should exploit any additional structure added to the problem. Instead of searching for feasible solutions on a straight line through the box (as the RFO does), the search can take place along any continuously elementwise-increasing curve between the lower and upper corner.
In addition, the feasibility problems can be simplified under perfect CSI and certain power constraints, as discussed in Section \ref{subsection_feasibility_problems}. The special case of weighted sum rate optimization and perfect CSI was recently considered in \cite{Eriksson2010a} and \cite{Weeraddana2010a}. Under a total power constraint, \cite{Eriksson2010a} formulated the search in a box as an approximate convex problem, which greatly improves the bounding step. With single-antenna transmitters, \cite{Weeraddana2010a} showed that the feasibility problems can be solved by simply checking the spectral radius of a matrix. Finally, note that all the system model extensions discussed in Section \ref{subsection_extensions} are possible for the BRB algorithm.
\end{remark}

\section{Low-complexity Suboptimal Strategies}

\label{section_low-complexity_strategies}

In general multicell systems, it is easy to derive feasible suboptimal transmission strategies, but very difficult to evaluate their performance. The BRB algorithm in Section \ref{section_monotonic_optimization} is suitable for computing benchmarks for such strategies. To evaluate the performance using some beamforming vectors $(\vect{v}_1,\ldots,\vect{v}_{K_r})$ and equalizing coefficients $(r_1,\ldots,r_{K_r})$, we need to know the robust MSE $\gamma_k$ that each $\textrm{MS}_k$ achieves (i.e., $\max_{\vect{h}_k \in \mathcal{U}_k} \text{MSE}_k \leq \gamma_k$).
With notation $\tilde{\gamma}_k= \sqrt{\gamma_k}$, the robust MSE $\gamma_k$ is easily obtained by solving
\begin{equation} \label{eq_evaluate_performance}
\begin{split}
\minimize{\tilde{\gamma}_k\geq 0, \lambda_k \geq 0}\,\, & \,\,\tilde{\gamma}_k \\
\mathrm{subject}\,\,\mathrm{to}\, & \,\,
\vect{A}_k \succeq \vect{0}_{N+K_r+2}
\end{split}
\end{equation}
which is a convex problem in $\tilde{\gamma}_k$ and $\lambda_k$. The matrix $\vect{A}_k$ is given in \eqref{eq_Ak-matrix} using $\tilde{r}_k=r_k^{-1}$ and $\tilde{\gamma}_k= \sqrt{\gamma_k}$. Next, we derive a low-complexity strategy based on interference constraints.

\subsection{Zero-Forcing and Interference-Constrained Beamforming}

A common way of simplifying transmission optimization is to constrain the co-user interference and to pretend that the channel estimates $\widehat{\vect{h}}_k$ are the actual channels. This approach is powerful both from a computational perspective and in terms of performance in the high SNR regime \cite{Jindal2006a}, but requires sufficiently many degrees of freedom (e.g., $|\mathcal{C}_j|\leq N_j$ for all $\textrm{BS}_j$). By pretending that the channel estimates are perfect, the "optimal" equalizing coefficient for $\textrm{MS}_k$ is $r_k=(\vect{v}_{k}^H \vect{D}_{k}^H \vect{C}_k^H \widehat{\vect{h}}_k) / (\sum_{\bar{k}} |\widehat{\vect{h}}_k^H \vect{C}_k \vect{D}_{\bar{k}} \vect{v}_{\bar{k}} |^2 + \sigma_k^2)$ and the "resulting" performance is
$\textrm{MSE}_k = (\sum_{\bar{k} \neq k} |\vect{h}_{k}^H \vect{C}_k \vect{D}_{\bar{k}} \vect{v}_{\bar{k}} |^2 + \sigma_k^2) / ( \sum_{\bar{k}} |\vect{h}_{k}^H \vect{C}_k \vect{D}_{\bar{k}} \vect{v}_{\bar{k}} |^2 + \sigma_k^2)$.

We consider $g_k(\textrm{MSE}_k) = 1/\textrm{MSE}_k -1$ to achieve the following tractable problem formulation:
\begin{equation} \label{eq_zero-forcing_original}
\begin{split}
\maximize{\substack{\vect{v}_{1},\ldots,\vect{v}_{K_r} \\ \eta_1,\ldots,\eta_{K_r} }   }\,\, & \,\, f(\eta_1,\ldots,\eta_{K_r}), \\
\mathrm{subject}\,\,\mathrm{to}\, & \,\,\, \frac{|\widehat{\vect{h}}_k^H \vect{C}_k \vect{D}_{k} \vect{v}_{k}|^2}{\sigma_k^2 + z_{k}} \geq \eta_k \quad \forall k, \\
& \,\,\, \sum_{\bar{k} \neq k} |\widehat{\vect{h}}_k^H \vect{C}_k \vect{D}_{\bar{k}} \vect{v}_{\bar{k}}|^2 \leq z_{k} \quad \forall k, \\ & \,\,\,
 \sum_{k} \vect{v}_k^H \vect{Q}_l \vect{v}_k \leq q_l \quad \forall l.
\end{split}
\end{equation}
The variables $\eta_k$ were introduced to clarify that this problem can be solved optimally if $f(\cdot)$ is concave.
If $z_{k}=0 \,\, \forall k$, \eqref{eq_zero-forcing_original} gives the \emph{zero-forcing beamforming} solution. Observe that
$z_{k}$ defines the maximal total interference that may be caused to $\textrm{MS}_k$, thus $z_{k}>0$ can be called \emph{interference-constrained beamforming}.
It could be suitable to select $z_{k}> 0$ under CSI uncertainty since the actual interference cannot nulled anyway. The following lemma shows that \eqref{eq_zero-forcing_original} can be solved as a convex optimization problem for any $z_{k} \geq 0 \,\, \forall k$.
\begin{lemma} \label{lemma_ZF_solution}
The semi-definite relaxation of \eqref{eq_zero-forcing_original} is
\begin{align} \label{eq_zero-forcing}
\maximize{\substack{ \vect{V}_{1}\succeq \vect{0},\ldots,\vect{V}_{K_r}\succeq \vect{0} \\ \eta_1,\ldots,\eta_{K_r} } }\,\, & \,\, f(\eta_1,\ldots,\eta_{K_r}), \\ \notag
\mathrm{subject}\,\,\mathrm{to}\, & \,\,\,
\frac{\trace\{\vect{D}_{k}^H \vect{C}_k^H \widehat{\vect{h}}_k \widehat{\vect{h}}_k^H \vect{C}_k \vect{D}_{k} \vect{V}_{k} \}}{\sigma_k^2+z_{k}} \geq \eta_k   \quad \forall k, \\ \notag
& \,\,\, \sum_{\bar{k} \neq k} \trace\{\vect{D}_{\bar{k}}^H \vect{C}_k^H \widehat{\vect{h}}_k \widehat{\vect{h}}_k^H \vect{C}_k \vect{D}_{\bar{k}} \vect{V}_{\bar{k}} \} \leq z_{k} \quad \forall k,\\ \notag & \,\,\,
 \sum_{k} \trace\{ \vect{Q}_l \vect{V}_{k} \} \leq q_l \quad \forall l.
\end{align}
This problem is convex if $f(\cdot)$ is a (strictly increasing) concave function. It always has rank-one solutions $\vect{V}^*_{k}=\vect{v}^*_{k} (\vect{v}^*_{k})^H$, where $\vect{v}^*_{k}$ solves the original problem in \eqref{eq_zero-forcing_original}.
\end{lemma}
\begin{IEEEproof}
The methodology in \cite{Bjornson2011a} and \cite{Wiesel2008a} can be used to show the existence of rank-one solutions. If an optimization procedure still delivers a high-rank solution $\vect{V}^*_{k}$, one can find $\vect{v}^*_{k}$ by maximizing $\Re\{ \widehat{\vect{h}}_k^H \vect{C}_k \vect{D}_{k} \vect{v}_{k} \}$ under the interference constraints
$|\widehat{\vect{h}}_{\bar{k}}^H \vect{C}_{\bar{k}} \vect{D}_{k} \vect{v}_{k}|^2 \leq \trace\{\vect{D}_{k}^H \vect{C}_{\bar{k}}^H \widehat{\vect{h}}_{\bar{k}} \widehat{\vect{h}}_{\bar{k}}^H \vect{C}_{\bar{k}} \vect{D}_{k} \vect{V}^*_{k} \}$  $\forall \bar{k} \neq k$ and power constraints $\vect{v}_k^H \vect{Q}_l \vect{v}_k \leq \trace\{ \vect{Q}_l \vect{V}^*_{k}\} \,\, \forall l$.
\end{IEEEproof}

Based on Lemma \ref{lemma_ZF_solution}, we can solve the zero-forcing problem for all concave system performance functions, for example, sum performance, proportional fairness, and max-min fairness. This can be viewed as an extension of the work in \cite{Wiesel2008a} to multicell systems with arbitrary power constraints. Zero-forcing and interference-constrained beamforming are suboptimal, but their simplicity make them good low-complexity strategies and \eqref{eq_zero-forcing_original} actually provides the optimal solution under perfect CSI if we happen to select the interference constraints optimally (see \cite{Rossi2011a} for a systematic method to search over $\{ z_k \}$). The performance under uncertain CSI is evaluated in Section \ref{section_numerical_examples}.
Observe that zero-forcing and interference-constrained beamforming can be used to achieve an initial lower bound for the BRB algorithm and thereby speed up the convergence.

\section{Numerical Examples}
\label{section_numerical_examples}

In this section, the robust monotonic optimization framework is evaluated numerically, using the YALMIP toolbox of \cite{YALMIP} and the numerical convex optimization solver SDPT3 from \cite{SDPT3}. First, the concept of robust performance regions and system performance functions are illustrated. Then, the performance of the zero-forcing approximation strategy in Section \ref{section_low-complexity_strategies} is compared with the optimal solution, calculated using the proposed BRB algorithm. Finally, the computational complexity of solving the RFO problem is exemplified and the convergence of the BRB algorithm is compared with the outer polyblock approximation algorithm.

While this section concentrates of illustrating different aspects of the proposed robust optimization framework, it is worth noting that our BRB algorithm was applied for performance benchmarking in realistic multicell systems with 20 users in \cite{Bjornson2011a}.

\begin{figure}
\subfigure[User Performance Measure: Inverse MSE, $g_k(\text{MSE}_k)=\text{MSE}^{-1}_k-1$.]
{\label{figure_region_MSE}\includegraphics[width=\columnwidth]{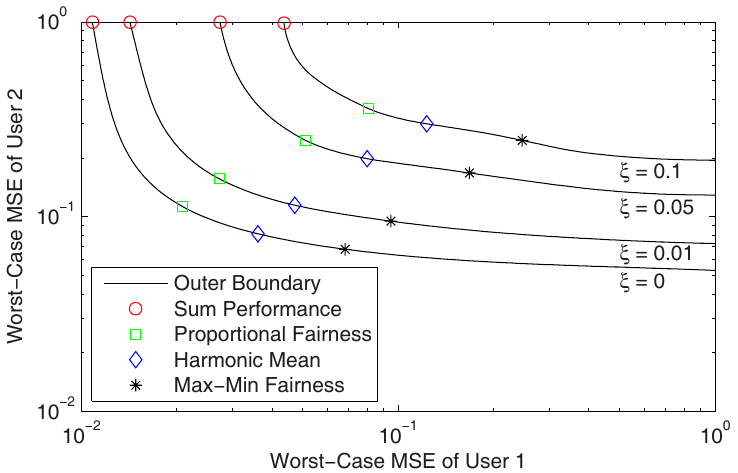}}\hfill
\subfigure[User Performance Measure: Data rate, $g_k(\text{MSE}_k)=\log_2(\text{MSE}^{-1}_k)$]
{\label{figure_region_rate}\includegraphics[width=\columnwidth]{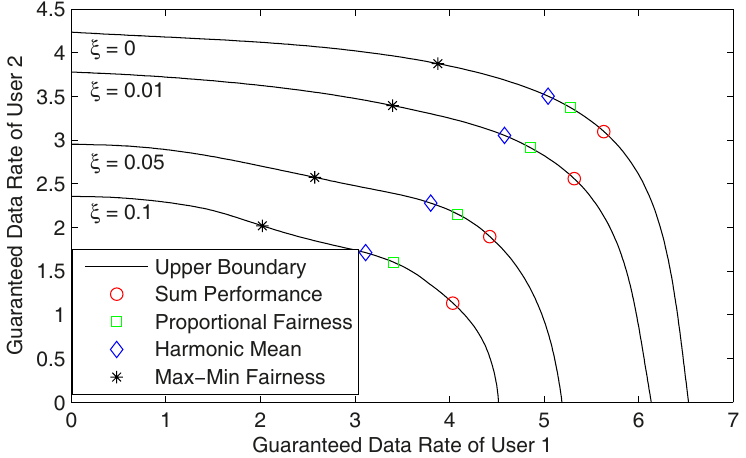}}
\caption{Robust performance regions with different squared radius $\xi$ of the channel uncertainty sets. The user performance measure is either (a) the inverse MSE or (b) the guaranteed data rate. The optimal points with different system performance functions are shown.} \label{figure_region}
\end{figure}

\subsection{Robust Performance Regions}

To illustrate the shape of the robust performance regions, we consider a simple network MIMO scenario with $K_r=2$ users. The total number of transmit antennas is $N=3$, all channels are Rayleigh fading and spatially uncorrelated, and we use per-antenna constraints with $q_l=10$ (i.e., 10 dB). The average SNR $\mathbb{E}\{ \| \vect{h}_{k} \|^2_2\}/\sigma^2_k$ is $N$ for user 1 and $N/4$ for user 2, creating an asymmetry that will highlight properties of different system performance functions. Spherical uncertainty sets $\mathcal{U}_k(\widehat{\vect{h}}_k,\vect{B}_{k})$ are assumed with $\vect{B}_{k}= \sqrt{\xi} \vect{I}_N$ in \eqref{eq_def_uncertainty_sets}, where the parameter $\sqrt{\xi}$ is the radius of the sphere. If one standard deviation of the channel estimation error is used as uncertainty set \cite{Bjornson2010a}, then $\xi$ equals the estimation error variance.

Fig.~\ref{figure_region} shows the robust performance regions for a random channel realization and different $\xi$. In Fig.~\ref{figure_region_MSE}, the inverse MSE is the user performance measure (i.e., $g_k(\text{MSE}_k)=\text{MSE}^{-1}_k-1$ to make $g_k(1)=0$), but the figure axes show MSEs to enhance viewing.
The guaranteed data rate $g_k(\text{MSE}_k)=\log_2(\text{MSE}^{-1}_k)$ is the user performance measure in Fig.~\ref{figure_region_rate}.
In both figures, the optimal system performance points are shown for the four measures exemplified in Section \ref{subsection_system_performance_measures}: sum performance, proportional fairness, harmonic mean, and max-min fairness.

Robustness towards channel uncertainty clearly decreases the size of the performance regions, without affecting the general shape (in this scenario). The optimal points of the four system performance measures are all on the upper boundaries (confirming Lemma \ref{lemma_global_optimum_attained_on_boundary}), but at quite different places. By introducing user weights in the system measures, the optimal points can be moved around on the upper boundary; in fact, the upper boundaries in Fig.~\ref{figure_region} were generated by solving max-min fairness optimization problems for a large set of weights (i.e., Algorithm 1 with different fairness-profiles).

\subsection{Evaluation of Zero-Forcing Beamforming}

Next, we evaluate the performance and robustness of the zero-forcing and interference-constrained beamforming approaches in Section \ref{section_low-complexity_strategies} (with maximal acceptable interference $z_k=0$ and $z_k>0$, respectively). The optimal solution, derived by the BRB algorithm, is used for benchmarking. We consider a scenario where $K_t=2$ base stations (with $N_j=3$ antennas each) jointly serve $K_r=6$ users. The channels $\vect{h}_{jk}$ are modeled as uncorrelated Rayleigh fading. The users are located such that
$\mathbb{E}\{ \| \vect{h}_{jk} \|^2_2\}/\sigma^2_k$ is $N_j$ for half the users and $N_j/2$ for the others, and vice versa for the other base station.
Per-base station constraints are considered with the power $q_l$ as a parameter that will be varied.
Proportional fairness of the user MSEs is used as system performance measure, thus the geometric mean MSE is minimized.\footnote{Under perfect CSI, minimizing the geometric mean MSE is identical to maximizing the sum rate. This equivalence does not hold under channel uncertainty, but a lower bound on the sum rate is maximized \cite{Shenouda2009a}.} This is achieved by having $g_k(\text{MSE}_k)=\text{MSE}^{-1}_k-1$ and $f(\vect{g})=\prod_k (g_k+1)^{1/K_r}$.

\begin{figure}
\includegraphics[width=\columnwidth]{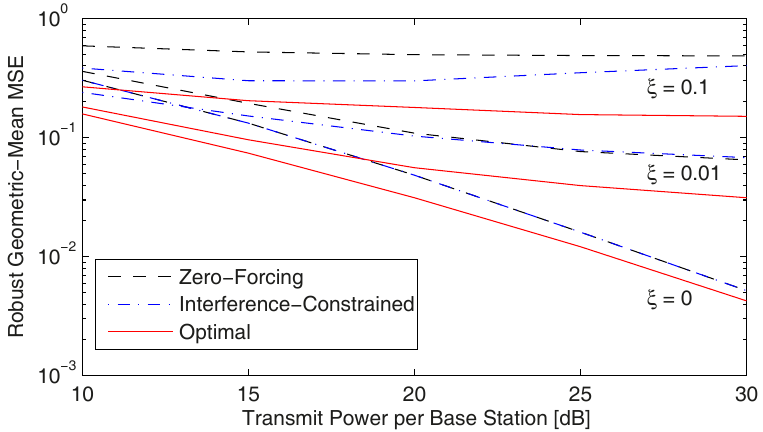} \vskip -2mm
\caption{Robust MSE with proportional fairness optimization (i.e., geometric mean). Zero-forcing and interference-constrained beamforming is compared with the optimal solution for different squared radius $\xi$ of the channel uncertainty sets.}\label{figure_ZF_comp} \vskip -3mm
\end{figure}

Fig.~\ref{figure_ZF_comp} shows the performance as a function of $q_l$ (the power per transmitter) and with spherical uncertainty sets with $\vect{B}_{k}= \sqrt{\xi} \vect{I}_N$. For small values on $\xi$ (i.e., low uncertainty), the zero-forcing approach provides close-to-optimal performance in the high SNR regime. At low SNR and for larger values on $\xi$, zero-forcing is clearly suboptimal. Interference-constrained beamforming with $z_k=(K_r-1) \tr\{ \vect{B}_{k} \vect{B}_{k}^H \}=(K_r-1) N\xi$ achieves better performance in these regimes, but the difference decreases with SNR which indicates that $z_k$ should be adapted to the SNR. In summary, zero-forcing is good at high SNR and robust to small channel uncertainties. By allowing interference $z_k>0$, interference-constrained beamforming achieves better performance at low SNR and larger uncertainties.

\subsection{Computational Complexity of RFO}

\begin{figure}
\begin{center}
\includegraphics[width=\columnwidth]{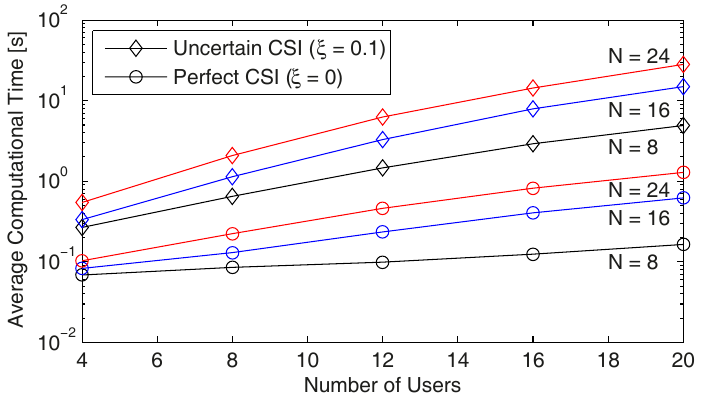} \vskip -2mm
\end{center}
\caption{Example of the average computational time of solving the feasibility problem in \eqref{eq_subproblem_performance-profile} with uncertain and perfect CSI. Different number of users and total number of antennas $N=\sum_j N_j$ are considered.}\label{figure_complexity} \vskip -3mm
\end{figure}

Next, the computational complexity of the robust fairness-profile optimization problem is evaluated. We consider the same multicell scenario as in the previous subsection, but we fix the transmit power per base station at 10 dB and vary the number of users $K_r \in \{4,8,12,16,20 \}$ and number of transmit antennas $N_1 = N_2 \in \{4,8,12\}$.
Recall that the RFO problem is solved by iterating the feasibility problem in \eqref{eq_subproblem_performance-profile} until a given line-search accuracy is achieved (10-15 iterations usually give a good accuracy, but it depends on the user performance functions). Therefore, Fig.~\ref{figure_complexity} shows the average computational time for solving this subproblem.
The simulation was performed at a standard PC running Linux/Ubuntu with an Intel Core2Duo Q8400 with 2.66 GHz using all four cores.

Fig.~\ref{figure_complexity} shows how the computational time increases with both the number of users and total number of antennas. It is clear that the complexity under channel uncertainty ($\xi=0.1$) is several times higher and has a steeper slope than under perfect CSI. This was expected from the discussion in Section \ref{subsection_feasibility_problems}, since the convex representation of the MSE constraints has larger dimensions under uncertain CSI. Comparing all the scenarios, it is clear that the computational time spans from a fraction of a second to a fraction of a minute; thus, the RFO problem can be solved quite efficiently (even at a standard PC with a general-purpose numerical solver) and is applicable for future real-time applications.

\subsection{Convergence Evaluation of BRB Algorithm}

In this subsection, the convergence of the proposed BRB algorithm is compared with the outer polyblock approximation algorithm in \cite{Brehmer2010a,Utschick2012a} and with a simple brute force approach\footnote{With \emph{brute force} we mean dividing the initial box of the BRB algorithm into (very many)  subboxes such that the difference between the lower and upper corner is less than $\varepsilon$ in each box. We solve one feasibility problem per box to find the optimum.}. As these algorithms are rather different, it is not meaningful to compare the number of iterations. Instead, we consider the performance as a function of the number of feasibility evaluations of the type in \eqref{eq_subproblem_performance-profile}. This convex subproblem is the main source of complexity in all the three approaches.

\begin{figure}
\begin{center}
\includegraphics[width=\columnwidth]{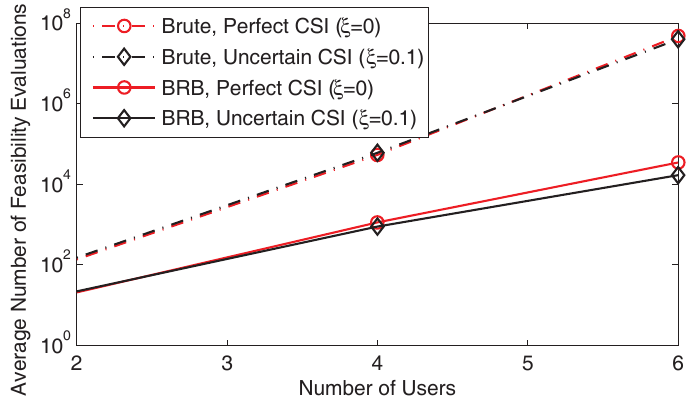} \vskip -2mm
\end{center}
\caption{Example of the average number of feasibility evaluations required to find the optimal solution (with a relative error of 0.1) with uncertain and perfect CSI.}\label{figure_brbcomplex} \vskip -3mm
\end{figure}

We first compare the BRB algorithm with brute force and consider the same multicell scenario as in the previous subsection. Fig.~\ref{figure_brbcomplex} shows the average number of feasibility evaluations (to achieve a relative error of 0.1) as a function of $K_r$. We let the number of antennas scale with the number of users as $N_j = K_r/2$ (we also have $\delta=1$). Fig.~\ref{figure_brbcomplex} reveals that the BRB algorithm, despite the unavoidable exponential complexity in $K_r$, is much more efficient than a brute force approach. We also observe that channel uncertainty slightly reduces the number of evaluations, basically since the performance region becomes smaller.

Finally, we compare the BRB algorithm with the outer polyblock approximation algorithm of \cite{Brehmer2010a,Utschick2012a} and consider a similar scenario as in these papers. Thus, we have $K_t=2$ transmitters with $N_1=N_2=3$ antennas and perfect CSI. Each transmitter serves two unique users (i.e., $K_r=4$), while coordinating interference to all users. The average SNR $\mathbb{E}\{ \| \vect{h}_{jk} \|^2_2\}/\sigma^2_k$ is $N_j$ if user $k \in \mathcal{D}_j$ and $N_j/2$ if $k \not \in \mathcal{D}_j$. Each transmitter has its own total power constraint with $q_j=10$ (i.e., 10 dB for single-user transmission) and the sum rate is chosen as performance measure.

\begin{figure}
\begin{center}
\includegraphics[width=\columnwidth]{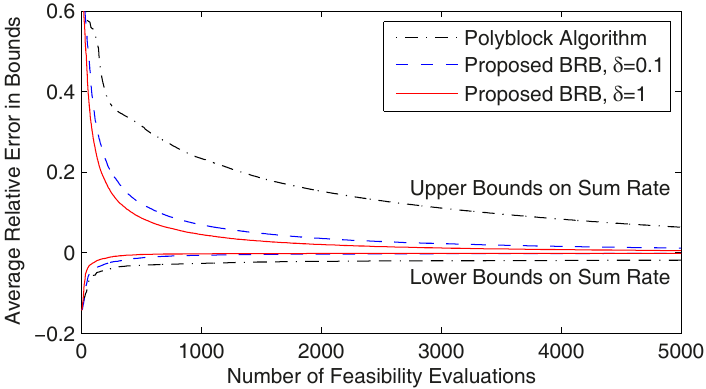}
\end{center} \vskip -2mm
\caption{Relative error of lower and upper bounds on the sum rate as a function of the number of feasibility evaluations.}\label{figure_sumrate_comparison} \vskip -3mm
\end{figure}

Fig.~\ref{figure_sumrate_comparison} shows the average relative deviations\footnote{If $f_{\text{opt}}$ is the optimal solution, the relative deviations of the lower and upper bound are $(f_{\min}-f_{\text{opt}})/f_{\text{opt}}$ and $(f_{\max}-f_{\text{opt}})/f_{\text{opt}}$, respectively.} (over 250 channel realizations) of the lower and upper bounds on the sum rate as a function of the number of feasibility evaluations. The BRB algorithm is used with a line-search accuracy $\delta$ of either $0.1$ or $1$, while $\delta=0.1$ was used for the polyblock algorithm.
Both algorithms quickly find feasible solutions within a few percent from the optimal value, but many evaluations are required to achieve a tight upper bound.
However, the proposed BRB algorithm shows much faster convergence in both the lower and the upper bound; after 5000 feasibility evaluations, the polyblock algorithm has still not reached the accuracy that the BRB algorithm achieved with 1000 evaluations. This is consistent with observations in \cite{Tuy2005a}, where the difference is also claimed to increase with the number of users.
Thus, in terms of achieving an $\varepsilon$-approximation on the optimal performance, the proposed BRB algorithm shows much faster convergence. For the BRB algorithm, $\delta=1$ gives faster convergence than $\delta=0.1$, indicating that having many iterations with loose bounds sometimes is more efficient than having few iterations with tight bounds.

\section{Conclusions}

This paper presented an optimization framework for resource allocation in multicell MISO downlink systems with general power constraints and robustness towards channel uncertainty. For any system performance measure that increases monotonically in each user's performance, the proposed \emph{branch-reduce-and-bound (BRB) algorithm} solves the resource allocation to global optimality. In each iteration, a line-search is performed in the robust performance region---a quasi-convex fairness-profile optimization problem that can be solved efficiently. Since most multiuser resource allocation problems are non-convex and NP-hard, the BRB algorithm is mainly suitable for computing benchmarks due to high computational complexity. The benchmarking capability was illustrated numerically by comparing it with a simple zero-forcing approximation. In addition, the BRB algorithm was shown to provide far better convergence than the previously known outer polyblock approximation algorithm.

\appendices

\section{Improvements under perfect CSI}
\label{appendix_perfect_CSI}

Both the robust fairness-profile optimization problem in Section \ref{section_opt_fairnessprofile} and the monotonic optimization framework in Section \ref{section_monotonic_optimization} can be directly applied to the case of perfect CSI (i.e., $\mathcal{U}_k=\{\widehat{\vect{h}}_k\}$). However, the computational complexity can be reduced by observing that optimal equalizing coefficients can be achieved by differentiation of $\textrm{MSE}_k$ in \eqref{eq_MSEk}: $r^{\textrm{optimal}}_k=( \vect{v}_{k}^H \vect{D}_{k}^H \vect{C}_k^H \vect{h}_{k}) / (\sum_{\bar{k}} |\vect{h}_{k}^H \vect{C}_k \vect{D}_{\bar{k}} \vect{v}_{\bar{k}} |^2 + \sigma_k^2)$. If this value is plugged into \eqref{eq_MSEk}, we achieve
\begin{equation}
\text{MSE}_k = \frac{\sum_{\bar{k} \neq k} |\vect{h}_{k}^H \vect{C}_k \vect{D}_{\bar{k}} \vect{v}_{\bar{k}} |^2 + \sigma_k^2}{ \sum_{\bar{k}} |\vect{h}_{k}^H \vect{C}_k \vect{D}_{\bar{k}} \vect{v}_{\bar{k}} |^2 + \sigma_k^2 }.
\end{equation}
Based on this MSE expression, the feasibility problem in \eqref{eq_subproblem_performance-profile} of Lemma \ref{lemma_fairness_profile_bisection} can be replaced by
\begin{equation} \label{eq_subproblem_performance-profile_perfect_CSI}
\begin{split}
\mathrm{find}\,\, & \,\, \vect{v}_1,\ldots,\vect{v}_{K_r} \\
\mathrm{subject}\,\,\mathrm{to}\,\, & \,\, \sum_{k} \vect{v}_k^H \vect{Q}_l \vect{v}_k \leq q_l \quad \forall l, \\
\sqrt{\frac{\gamma_k}{1-\gamma_k}} \vect{h}_{k}^H \vect{C}_k &\vect{D}_{k} \vect{v}_{k} \geq \sqrt{ \sum_{\bar{k}} |\vect{h}_{k}^H \vect{C}_k \vect{D}_{\bar{k}} \vect{v}_{\bar{k}} |^2 + \sigma_k^2} \,\, \forall k.
\end{split}
\end{equation}
This new feasibility problem is also convex, because the MSE constraints have been turned into second-order cone constraints (see \cite{Bengtsson2001a,Wiesel2006a,Yu2007a} for details on this approach).
However, \eqref{eq_subproblem_performance-profile_perfect_CSI} has fewer optimization variables which means lower computational complexity.
Thus, \eqref{eq_subproblem_performance-profile_perfect_CSI} should always be used instead of \eqref{eq_subproblem_performance-profile} under perfect CSI.

\section{}

\subsubsection*{\textbf{Proof of Lemma \ref{lemma_performance_region_normal}}}
\label{appendix_performance_region_normal}

To prove that the set is normal, take $\vect{x}=(x_1,\ldots,x_{K_r}) \in \mathcal{R}$ and assume that $\{r_k^*\}_{k=1}^{K_r}$ and $\{\vect{v}_k^*\}_{k=1}^{K_r}$ is a feasible solution that attains this point. We want to show that any $\vect{x}'= (x'_1,\ldots,x'_{K_r}) \in \mathbb{R}_+^{K_r}$ with $\vect{x}' \leq \vect{x}$ also belongs to $\mathcal{R}$. To this end, we fix the beamforming vectors at $\{\vect{v}_k^*\}_{k=1}^{K_r}$ and search for equalizing coefficients $\{r_{\vect{x}',k}\}_{k=1}^{K_r}$ that gives $\widetilde{\text{MSE}}_k = \gamma'_k$ for all $k$, where $\gamma'_k=g_k^{-1} (x'_k)$. For a given channel realization $\vect{h}_k$, denote the MSE in \eqref{eq_MSEk} as $\text{MSE}_k(r_k,\vect{h}_k)$. Observe that $\text{MSE}_k(r_k,\vect{h}_k)$ is a second-order polynomial in $r_k$ that has a unique minimum and then approaches infinity continuously as $r_k \rightarrow \infty$. Thus, we can solve the second-order equation $\text{MSE}_k(r_k,\vect{h}_k)=\gamma'_k$ to derive the largest root
\begin{equation}
r'_k(\vect{h}_k) = \frac{a_k+\sqrt{a_k^2-(1-\gamma'_k)b_k}}{b_k}
\end{equation}
where $a_k=\Re(\vect{h}_{k}^H \vect{C}_k \vect{D}_{k} \vect{v}_k)$ and $b_k=\| \vect{h}_{k}^H \vect{C}_k [\vect{D}_{1} \vect{v}_1 \,\ldots\,  \vect{D}_{K_r} \vect{v}_{K_r}]\|_2^2+\sigma_k^2$. This solution will be real-valued, because it is real-valued for $\vect{x}' = \vect{x}$ and increases with $\gamma'_k$. Since we consider the largest root, $\textrm{MSE}_k(r_k,\vect{h}_k)>\gamma'_k$ for all $r_k > r'_k(\vect{h}_k)$. By selecting
\begin{equation}
r_{\vect{x}',k} = \min_{\vect{h}_k \in \mathcal{U}_k} r'_k(\vect{h}_k) \quad \forall k
\end{equation}
we can make sure that $g_k(\widetilde{\text{MSE}}_k) = x'_k$ and thus that $\vect{x}' \in \mathcal{R}$.

Next, we prove that $\mathcal{R}$ is a compact set. First, observe that the set of feasible beamforming vectors, $\mathcal{V}$, in \eqref{eq_feasible_transmit_strategies} is compact.
Next, observe that it is sufficient to search for equalizing coefficients $r_k$ in the compact set $\mathcal{E}_k = [0,1/\sigma_k]$, since greater values make the noise part of the MSE in \eqref{eq_MSEk} larger than one (and thus, $\text{MSE}_k \geq 1$). The MSEs are continuous functions of the beamforming vectors and equalizing coefficients, and the performance functions $g_k(\widetilde{\text{MSE}}_k)$ are continuous by definition. Therefore, $g_k( \min (\max_{\vect{h}_k \in \mathcal{U}_k} \text{MSE}_k,1))$ is continuous for any compact set $\mathcal{U}_k$. Finally, we invoke \cite[Theorem 4.14]{Rudin1976a}, which says that the continuous mapping of a compact set is also a compact set. Since $\mathcal{R}$ is the image of a continuous mapping from $\mathcal{V}$ and $\mathcal{E}_k$, the robust performance region is compact.

\subsubsection*{\textbf{Proof of Theorem \ref{theorem_convex_feasibility_problem}}}
\label{appendix_robust_reformulation}

The proof is based on the following well-known result in robust worst-case optimization theory:

\begin{lemma} \label{lemma_modified_s-lemma}
Given $\vect{A}, \vect{P}, \vect{Q}$, with $\vect{A}=\vect{A}^H$, the expression
\begin{equation} \label{eq_lemma_robust_LMI_first}
\vect{A} \succeq \vect{P}^H \vect{Z} \vect{Q} + \vect{Q}^H \vect{Z}^H \vect{P} \quad \quad \forall \vect{Z}: \| \vect{Z} \|_2 \leq \varrho
\end{equation}
holds if and only if
\begin{equation} \label{eq_lemma_robust_LMI_second}
\exists \lambda \in \mathbb{R}_+ \quad \text{s.t.}  \,\, \left[\begin{IEEEeqnarraybox*}[][c]{,c/c,}
\vect{A}-\lambda \vect{Q}^H \vect{Q} & -\varrho \vect{P}^H \\
-\varrho \vect{P} & \lambda \vect{I}%
\end{IEEEeqnarraybox*}  \right] \succeq \vect{0}.
\end{equation}
\end{lemma}
\begin{IEEEproof}
The proof is given in \cite[Proposition 2]{Eldar2004a}.
\end{IEEEproof}

Lemma \ref{lemma_modified_s-lemma} can be used to reformulate each MSE constraint
\begin{equation} \label{eq_MSE-constraint_original}
\max_{\vect{h}_k \in \mathcal{U}_k} \text{MSE}_k \leq \gamma_k
\end{equation}
in \eqref{eq_subproblem_performance-profile} as a semi-definite constraint. First, we replace $\vect{h}_k$ with $\widehat{\vect{h}}_k+\vect{B}_k \tilde{\boldsymbol{\epsilon}}_{k}$ in the MSE expression of \eqref{eq_MSEk}. Next, we apply Schur complement lemma \cite[Theorem 1.12]{Zhang2005a} to rewrite \eqref{eq_MSE-constraint_original} as
\begin{equation} \label{eq_MSE-constraint_sdp-version}
\begin{split}
&\left[\begin{IEEEeqnarraybox*}[][c]{c,c,c}
\sqrt{\gamma_k} r_k^{-1} & \widehat{\vect{h}}_k^H \vect{C}_k \widetilde{\vect{V}} \!-\!r_k^{-1} \vect{e}_k^T & \sigma_k\\
\widetilde{\vect{V}}^H \vect{C}_k^H \widehat{\vect{h}}_k \!-\!r_k^{-1} \vect{e}_k & \sqrt{\gamma_k} r_k^{-1} \vect{I}_{K_r} & \vect{0} \\
\sigma_k & \vect{0} & \sqrt{\gamma_k} r_k^{-1}%
\end{IEEEeqnarraybox*}  \right] \\ &\,\,\,
+ \left[\begin{IEEEeqnarraybox*}[][c]{c,c,c}
0 & \tilde{\boldsymbol{\epsilon}}_{k}^H \vect{B}_k^H \vect{C}_k \widetilde{\vect{V}} & 0\\
\widetilde{\vect{V}}^H \vect{C}_k^H \vect{B}_k \tilde{\boldsymbol{\epsilon}}_{k}  & \vect{0}_{K_r} & \vect{0} \\
0 & \vect{0} & 0%
\end{IEEEeqnarraybox*}  \right]
\succeq \vect{0} \quad \forall \tilde{\boldsymbol{\epsilon}}_{k} : \| \tilde{\boldsymbol{\epsilon}} \|_2 \leq 1.
\end{split}
\end{equation}
Finally, we apply Lemma \ref{lemma_modified_s-lemma} with $\vect{A}$ being the first matrix in \eqref{eq_MSE-constraint_sdp-version}, $\vect{P}=[\vect{0} \,\,\, \vect{B}_k^H \vect{C}_k \widetilde{\vect{V}} \, \,\, \vect{0}]$, $\vect{Q}=[-1 \,\,\, \vect{0} \,\,\, 0]$, $\vect{Z}=\tilde{\boldsymbol{\epsilon}}_{k}$, and $\varrho=1$. The obtained reformulation of \eqref{eq_MSE-constraint_sdp-version} is the constraint in \eqref{eq_expressed_as_sdp}. By optimizing over $\tilde{r}_k=r_k^{-1}$ instead of $r_k$, we observe that this constraint is linear in $\vect{v}_1,\ldots,\vect{v}_{K_r},\tilde{r}_k$. Thus, the reformulated problem is convex.
Further details are available in similar proofs (under different system assumptions) in for instance \cite{Bental2009a,Zheng2008b,Vucic2009a,Shenouda2009a}.

\subsubsection*{\textbf{Proof of Lemma \ref{lemma_reduction_algorithm}}}
\label{appendix_reduction_algorithm}

First, consider the reduction of the box from $[\tilde{\vect{a}}_l,\tilde{\vect{b}}_l]$ to $[\tilde{\vect{a}}'_l,\tilde{\vect{b}}_l]$ (i.e., from below). If the boxes are identical, no solutions are lost and we are finished. Otherwise, $\tilde{\vect{a}}_l \leq \tilde{\vect{a}}'_l$ with strict inequality in at least one element. For elements with strict inequality we have $\nu_k<1$, while $\nu_k=1$ holds for all other elements. There exist $\vect{g} \in [\tilde{\vect{a}}_l,\tilde{\vect{b}}_l]$ such that $\vect{g} \not \in [\tilde{\vect{a}}'_l,\tilde{\vect{b}}_l]$. For any such $\vect{g}$ there is a dimension $k$ such that $g_k < \tilde{a}'_{l,k}$ and $\nu_k<1$. Thus, $\vect{g} \leq \tilde{\vect{b}}_l - \tilde{\nu} (\tilde{b}_{l,k}-\tilde{a}_{l,k})\vect{e}_k$ for some $\tilde{\nu}$ with $\nu_k < \tilde{\nu} \leq 1$. For the system performance function, we have
\begin{equation} \label{eq_reduction_inequality}
\begin{split}
f(\vect{g}) &\leq f( \tilde{\vect{b}}_l - \tilde{\nu} (\tilde{b}_{l,k}-\tilde{a}_{l,k}) \vect{e}_k ) \\ &
< f( \tilde{\vect{b}}_l - \nu_k (\tilde{b}_{l,k}-\tilde{a}_{l,k}) \vect{e}_k ) = f_{\min}.
\end{split}
\end{equation}
The strict inequality follows since $\nu_k$ is selected to be the largest value that gives equality in the set defined in \eqref{eq_reduction_variables}. From \eqref{eq_reduction_inequality} it is clear that any $\vect{g}$ removed in the reduction (from below) will have a function value strictly below $f_{\min}$.
The reduction from above is proved analogously. Finally, observe that $[\tilde{\vect{a}}'_l,\tilde{\vect{b}}'_l] \subseteq [\tilde{\vect{a}}_l,\tilde{\vect{b}}_l]$ since each element in $\tilde{\vect{a}}'_l$ and $\tilde{\vect{b}}'_l$ are calculated as convex combinations of $\tilde{\vect{a}}_l$ and $\tilde{\vect{b}}_l$.

\subsubsection*{\textbf{Proof of Lemma \ref{lemma_bounding_algorithm}}}
\label{appendix_bounding_algorithm}

The line-search procedure in Algorithm 1 finds the best feasible solution (with accuracy $\delta$) on the line segment between $\vect{a}$ and $\vect{b}$. Thus, $\vect{a}+ \boldsymbol{\alpha} f^{\min}_{\text{RFO}}$ is feasible and can be used for a lower bound on the performance. Similarly, $\vect{n}$ is either on the upper boundary or infeasible. Since $\mathcal{R}$ is normal, there are no feasible points $\vect{x} \in \mathcal{M}$ with $\vect{x}>\vect{n}$. The corner points where all element but one are larger than in $\vect{n}$ are $ \vect{b} - (b_{k} - n_k) \vect{e}_k$ for $k=1,\ldots,K_r$. These can be used to calculate an upper bound on the performance.

\section*{Acknowledgement}

The first author would like to thank S. J\"{a}rmyr for stimulating discussions on performance optimization.

\bibliographystyle{IEEEtran}
\bibliography{IEEEabrv,refs}

\begin{thebibliography}{10}
\providecommand{\url}[1]{#1}
\csname url@samestyle\endcsname
\providecommand{\newblock}{\relax}
\providecommand{\bibinfo}[2]{#2}
\providecommand{\BIBentrySTDinterwordspacing}{\spaceskip=0pt\relax}
\providecommand{\BIBentryALTinterwordstretchfactor}{4}
\providecommand{\BIBentryALTinterwordspacing}{\spaceskip=\fontdimen2\font plus
\BIBentryALTinterwordstretchfactor\fontdimen3\font minus
  \fontdimen4\font\relax}
\providecommand{\BIBforeignlanguage}[2]{{%
\expandafter\ifx\csname l@#1\endcsname\relax
\typeout{** WARNING: IEEEtran.bst: No hyphenation pattern has been}%
\typeout{** loaded for the language `#1'. Using the pattern for}%
\typeout{** the default language instead.}%
\else
\language=\csname l@#1\endcsname
\fi
#2}}
\providecommand{\BIBdecl}{\relax}
\BIBdecl

\bibitem{Viswanath2003a}
P.~Viswanath and D.~Tse, ``Sum capacity of the vector {Gaussian} broadcast
  channel and uplink-downlink duality,'' \emph{{IEEE} Trans. Inf. Theory},
  vol.~49, no.~8, pp. 1912--1921, 2003.

\bibitem{Dohler2011a}
M.~Dohler, R.~Heath, A.~Lozano, C.~Papadias, and R.~Valenzuela, ``Is the {PHY}
  layer dead?'' \emph{{IEEE} Commun. Mag.}, vol.~49, no.~4, pp. 159--165, 2011.

\bibitem{Huh2010a}
H.~Huh, G.~Caire, S.-H. Moon, and I.~Lee, ``Multi-cell mimo downlink with
  fairness criteria: The large system limit,'' in \emph{Proc.~ISIT'10}, 2010,
  pp. 2058--2062.

\bibitem{SDPT3}
R.~T{\"u}t{\"u}nc{\"u}, K.~Toh, and M.~Todd, ``Solving
  semidefinite-quadratic-linear programs using {SDPT3},'' \emph{Mathematical
  Programming}, vol.~95, no.~2, pp. 189--217, 2003.

\bibitem{Bengtsson2001a}
M.~Bengtsson and B.~Ottersten, ``Optimal and suboptimal transmit beamforming,''
  in \emph{Handbook of Antennas in Wireless Communications}, L.~C. Godara,
  Ed.\hskip 1em plus 0.5em minus 0.4em\relax CRC Press, 2001.

\bibitem{Bjornson2011a}
E.~Bj{\"{o}}rnson, N.~Jald{\'e}n, M.~Bengtsson, and B.~Ottersten, ``Optimality
  properties, distributed strategies, and measurement-based evaluation of
  coordinated multicell {OFDMA} transmission,'' \emph{{IEEE} Trans. Signal
  Process.}, vol.~59, no.~12, pp. 6086--6101, 2011.

\bibitem{Wiesel2006a}
A.~Wiesel, Y.~Eldar, and S.~Shamai, ``Linear precoding via conic optimization
  for fixed {MIMO} receivers,'' \emph{{IEEE} Trans. Signal Process.}, vol.~54,
  no.~1, pp. 161--176, 2006.

\bibitem{Yu2007a}
W.~Yu and T.~Lan, ``Transmitter optimization for the multi-antenna downlink
  with per-antenna power constraints,'' \emph{{IEEE} Trans. Signal Process.},
  vol.~55, no.~6, pp. 2646--2660, 2007.

\bibitem{Schubert2004a}
M.~Schubert and H.~Boche, ``Solution of the multiuser downlink beamforming
  problem with individual {SINR} constraints,'' \emph{{IEEE} Trans. Veh.
  Technol.}, vol.~53, no.~1, pp. 18--28, 2004.

\bibitem{Schubert2005a}
------, ``{QoS}-based resource allocation and transceiver optimization,''
  \emph{Foundations and Trends in Communication and Information Theory},
  vol.~2, no.~6, pp. 383--529, 2005.

\bibitem{Bental2009a}
A.~Ben-Tal, L.~E. Ghaoui, and A.~Nemirovski, \emph{Robust Optimization}.\hskip
  1em plus 0.5em minus 0.4em\relax Princeton University Press, 2009.

\bibitem{Zheng2008b}
G.~Zheng, K.-K. Wong, and T.-S. Ng, ``Robust linear {MIMO} in the downlink: A
  worst-case optimization with ellipsoidal uncertainty regions,'' \emph{EURASIP
  J. on Adv. in Signal Process.}, 2008.

\bibitem{Vucic2009a}
N.~Vu$\mathrm{\check{c}i\acute{c}}$ and H.~Boche, ``Robust {QoS}-constrained
  optimization of downlink multiuser {MISO} systems,'' \emph{{IEEE} Trans.
  Signal Process.}, vol.~57, no.~2, pp. 714--725, 2009.

\bibitem{Shenouda2009a}
M.~B. Shenouda and T.~Davidson, ``Nonlinear and linear broadcasting with {QoS}
  requirements: Tractable approaches for bounded channel uncertainties,''
  \emph{{IEEE} Trans. Signal Process.}, vol.~57, no.~5, pp. 1936--1947, 2009.

\bibitem{Vucic2009b}
N.~Vu$\mathrm{\check{c}i\acute{c}}$, H.~Boche, and S.~Shi, ``Robust transceiver
  optimization in downlink multiuser {MIMO} systems,'' \emph{{IEEE} Trans.
  Signal Process.}, vol.~57, no.~9, pp. 3576--3587, 2009.

\bibitem{Tajer2011a}
A.~Tajer, N.~Prasad, and X.~Wang, ``Robust linear precoder design for
  multi-cell downlink transmission,'' \emph{{IEEE} Trans. Signal Process.},
  vol.~59, no.~1, pp. 235--251, 2011.

\bibitem{Chalise2007a}
B.~Chalise, S.~Shahbazpanahi, A.~Czylwik, and A.~Gershman, ``Robust downlink
  beamforming based on outage probability specifications,'' \emph{{IEEE} Trans.
  Wireless Commun.}, vol.~6, no.~10, pp. 3498--3503, 2007.

\bibitem{Shenouda2008a}
M.~B. Shenouda and T.~Davidson, ``Probabilistically-constrained approaches to
  the design of the multiple antenna downlink,'' in \emph{Proc.~Asilomar'08},
  2008, pp. 1120--1124.

\bibitem{Wang2011a}
K.-Y. Wang, T.-H. Chang, and C.-Y.~C. W.-K.~Ma, A.~So, ``Probabilistic {SINR}
  constrained robust transmit beamforming: A {Bernstein}-type inequality based
  conservative approach,'' in \emph{Proc.~IEEE ICASSP'11}, 2011, pp.
  3080--3083.

\bibitem{Liu2011a}
Y.-F. Liu, Y.-H. Dai, and Z.-Q. Luo, ``Coordinated beamforming for {MISO}
  interference channel: Complexity analysis and efficient algorithms,''
  \emph{{IEEE} Trans. Signal Process.}, vol.~59, no.~3, pp. 1142--1157, 2011.

\bibitem{Tuy2000a}
H.~Tuy, ``Monotonic optimization: Problems and solution approaches,''
  \emph{SIAM J. Optim.}, vol.~11, no.~2, pp. 464--494, 2000.

\bibitem{Tuy2005a}
H.~Tuy, F.~Al-Khayyal, and P.~Thach, ``Monotonic optimization: Branch and cut
  methods,'' in \emph{Essays and Surveys in Global Optimization}, C.~Audet,
  P.~Hansen, and G.~Savard, Eds.\hskip 1em plus 0.5em minus 0.4em\relax
  Springer US, 2005.

\bibitem{Brehmer2009b}
J.~Brehmer and W.~Utschick, ``Nonconcave utility maximisation in the {MIMO}
  broadcast channel,'' \emph{EURASIP J. on Adv. in Signal Process.}, 2009.

\bibitem{Brehmer2009a}
------, ``Utility maximization in the multi-user {MISO} downlink with linear
  precoding,'' in \emph{Proc.~IEEE ICC'09}, 2009.

\bibitem{Jorswieck2010a}
E.~Jorswieck and E.~Larsson, ``Monotonic optimization framework for the
  two-user {MISO} interference channel,'' \emph{{IEEE} Trans. Commun.},
  vol.~58, no.~7, pp. 2159--2168, 2010.

\bibitem{Brehmer2010a}
J.~Brehmer and W.~Utschick, ``Optimal interference management in multi-antenna,
  multi-cell systems,'' in \emph{Proc.~Int. Zurich Seminar on Commun.}, 2010,
  pp. 134--137.

\bibitem{Utschick2012a}
W.~Utschick and J.~Brehmer, ``Monotonic optimization framework for coordinated
  beamforming in multicell networks,'' \emph{{IEEE} Trans. Signal Process.},
  vol.~60, no.~4, pp. 1899--1909, 2012.

\bibitem{Zhang2010a}
R.~Zhang and S.~Cui, ``Cooperative interference management with {MISO}
  beamforming,'' \emph{{IEEE} Trans. Signal Process.}, vol.~58, no.~10, pp.
  5450--5458, 2010.

\bibitem{Karipidis2010a}
E.~Karipidis and E.~Larsson, ``Efficient computation of the {Pareto} boundary
  for the {MISO} interference channel with perfect {CSI},'' in
  \emph{Proc.~WiOpt'10}, 2010, pp. 573--577.

\bibitem{Bjornson2010d}
E.~Bj{\"{o}}rnson, M.~Bengtsson, and B.~Ottersten, ``Optimality properties and
  low-complexity solutions to coordinated multicell transmission,'' in
  \emph{Proc.~IEEE GLOBECOM'10}, 2010.

\bibitem{Bjornson2011d}
E.~Bj{\"{o}}rnson, M.~Bengtsson, G.~Zheng, and B.~Ottersten, ``Computational
  framework for optimal robust beamforming in coordinated multicell systems,''
  in \emph{Proc.~IEEE CAMSAP'11}, 2011.

\bibitem{Mochaourab2011a}
R.~Mochaourab and E.~Jorswieck, ``Optimal beamforming in interference networks
  with perfect local channel information,'' \emph{{IEEE} Trans. Signal
  Process.}, vol.~59, no.~3, pp. 1128--1141, 2011.

\bibitem{Shang2010a}
X.~Shang, B.~Chen, and H.~V. Poor, ``Multiuser {MISO} interference channels
  with single-user detection: Optimality of beamforming and the achievable rate
  region,'' \emph{{IEEE} Trans. Inf. Theory}, vol.~57, no.~7, pp. 4255--4273,
  2011.

\bibitem{Song2011a}
E.~Song, Q.~Shi, M.~Sanjabi, R.~Sun, and Z.-Q. Luo, ``Robust {SINR}-constrained
  {MISO} downlink beamforming: When is semidefinite programming relaxation
  tight?'' in \emph{Proc.~IEEE ICASSP'11}, 2011.

\bibitem{Shang2009b}
X.~Shang, G.~Kramer, B.~Chen, and H.~V. Poor, ``A new outer bound and the
  noisy-interference sum-rate capacity for {Gaussian} interference channels,''
  \emph{{IEEE} Trans. Inf. Theory}, vol.~55, no.~2, pp. 689--699, 2009.

\bibitem{Zhang2008a}
H.~Zhang, N.~Mehta, A.~Molisch, J.~Zhang, and H.~Dai, ``Asynchronous
  interference mitigation in cooperative base station systems,'' \emph{{IEEE}
  Trans. Wireless Commun.}, vol.~7, no.~1, pp. 155--165, 2008.

\bibitem{Bjornson2010a}
E.~Bj{\"{o}}rnson and B.~Ottersten, ``A framework for training-based estimation
  in arbitrarily correlated {Rician} {MIMO} channels with {Rician}
  disturbance,'' \emph{{IEEE} Trans. Signal Process.}, vol.~58, no.~3, pp.
  1807--1820, 2010.

\bibitem{Gershman2010a}
A.~Gershman, N.~Sidiropoulos, S.~Shahbazpanahi, M.~Bengtsson, and B.~Ottersten,
  ``Convex optimization-based beamforming,'' \emph{{IEEE} Signal Process.
  Mag.}, vol.~27, no.~3, pp. 62--75, 2010.

\bibitem{Karakayali2006a}
M.~Karakayali, G.~Foschini, and R.~Valenzuela, ``Network coordination for
  spectrally efficient communications in cellular systems,'' \emph{{IEEE}
  Wireless Commun. Mag.}, vol.~13, no.~4, pp. 56--61, 2006.

\bibitem{Jorswieck2008b}
E.~Jorswieck, E.~Larsson, and D.~Danev, ``Complete characterization of the
  {Pareto} boundary for the {MISO} interference channel,'' \emph{{IEEE} Trans.
  Signal Process.}, vol.~56, no.~10, pp. 5292--5296, 2008.

\bibitem{Palomar2003a}
D.~Palomar, J.~Cioffi, and M.~Lagunas, ``Joint {Tx}-{Rx} beamforming design for
  multicarrier {MIMO} channels: a unified framework for convex optimization,''
  \emph{{IEEE} Trans. Signal Process.}, vol.~51, no.~9, pp. 2381--2401, 2003.

\bibitem{Luo2008a}
Z.-Q. Luo and S.~Zhang, ``Dynamic spectrum management: Complexity and
  duality,'' \emph{{IEEE} J. Sel. Topics Signal Process.}, vol.~2, no.~1, pp.
  57--73, 2008.

\bibitem{Zhang2010b}
R.~Zhang and L.~Hanzo, ``Joint and distributed linear precoding for centralised
  and decentralised multicell processing,'' in \emph{Proc.~IEEE VTC'10-Fall},
  2010.

\bibitem{Mohseni2006a}
M.~Mohseni, R.~Zhang, and J.~Cioffi, ``Optimized transmission for fading
  multiple-access and broadcast channels with multiple antennas,'' \emph{{IEEE}
  J. Sel. Areas Commun.}, vol.~24, no.~8, pp. 1627--1639, 2006.

\bibitem{Maddah2009a}
M.~Maddah-Ali, A.~Mobasher, and A.~Khandani, ``Fairness in multiuser systems
  with polymatroid capacity region,'' \emph{{IEEE} Trans. Inf. Theory},
  vol.~55, no.~5, pp. 2128--2138, 2009.

\bibitem{Boyd1993a}
S.~Boyd and L.~E. Ghaoui, ``Method of centers for minimizing generalized
  eigenvalues,'' \emph{Linear Algebra and Its Applications}, vol. 188, pp.
  63--111, 1993.

\bibitem{Bental2001a}
A.~Ben-Tal and A.~Nemirovski, \emph{Lectures on modern convex optimization:
  analysis, algorithms, and engineering applications}.\hskip 1em plus 0.5em
  minus 0.4em\relax SIAM, 2001.

\bibitem{Scutari2008a}
G.~Scutari, D.~P. Palomar, and S.~Barbarossa, ``Cognitive {MIMO} radio,''
  \emph{{IEEE} Signal Process. Mag.}, vol.~25, no.~6, pp. 46--59, 2008.

\bibitem{Balakrishnan1991a}
V.~Balakrishnan, S.~Boyd, and S.~Balemi, ``Robust downlink beamforming based on
  outage probability specifications,'' \emph{Int. J. Robust and Nonlinear
  Control}, vol.~1, no.~4, pp. 295--317, 1991.

\bibitem{Eriksson2010a}
K.~Eriksson, S.~Shi, N.~Vu$\mathrm{\check{c}i\acute{c}}$, M.~Schubert, and
  E.~Larsson, ``Globally optimal resource allocation for achieving maximum
  weighted sum rate,'' in \emph{Proc.~GLOBECOM'10}, 2010.

\bibitem{Weeraddana2010a}
P.~Weeraddana, M.~Codreanu, M.~Latva-aho, and A.~Ephremides, ``Weighted
  sum-rate maximization for a set of interfering links via branch and bound,''
  in \emph{Proc.~Asilomar'10}, 2010.

\bibitem{Jindal2006a}
N.~Jindal, ``{MIMO} broadcast channels with finite-rate feedback,''
  \emph{{IEEE} Trans. Inf. Theory}, vol.~52, no.~11, pp. 5045--5060, 2006.

\bibitem{Wiesel2008a}
A.~Wiesel, Y.~Eldar, and S.~Shamai, ``Zero-forcing precoding and generalized
  inverses,'' \emph{{IEEE} Trans. Signal Process.}, vol.~56, no.~9, pp.
  4409--4418, 2008.

\bibitem{Rossi2011a}
M.~Rossi, A.~Tulino, O.~Simeone, and A.~Haimovich, ``Non-convex utility
  maximization in {Gaussian} {MISO} broadcast and interference channels,'' in
  \emph{Proc.~IEEE ICASSP'11}, 2011, pp. 2960--2963.

\bibitem{YALMIP}
\BIBentryALTinterwordspacing
J.~L{\"{o}}fberg, ``{YALMIP}: A toolbox for modeling and optimization in
  {MATLAB},'' in \emph{Proc.~IEEE CACSD}, 2004, pp. 284--289. [Online].
  Available: \url{http://users.isy.liu.se/johanl/yalmip}
\BIBentrySTDinterwordspacing

\bibitem{Rudin1976a}
W.~Rudin, \emph{Principles of Mathematical Analysis}.\hskip 1em plus 0.5em
  minus 0.4em\relax McGraw-Hill, 1976.

\bibitem{Eldar2004a}
Y.~Eldar and N.~Merhav, ``A competitive minimax approach to robust estimation
  of random parameters,'' \emph{{IEEE} Trans. Signal Process.}, vol.~52, no.~7,
  pp. 1931--1946, 2004.

\bibitem{Zhang2005a}
F.~Zhang, \emph{The {Schur} Complement and Its Applications}.\hskip 1em plus
  0.5em minus 0.4em\relax Springer, 2005.

\end{thebibliography}

\begin{IEEEbiography}[{\includegraphics[width=1in,height=1.25in,clip,keepaspectratio]{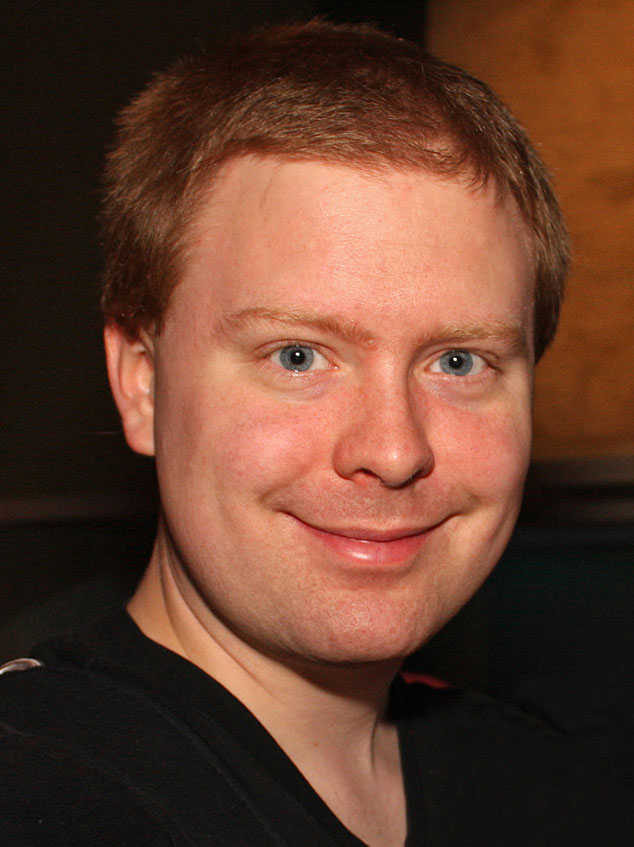}}]{Emil Bj\"ornson}
(S'07-M'12) was born in Malm\"o, Sweden, in 1983. He received the M.S. degree in Engineering Mathematics from Lund University, Lund, Sweden, in 2007. He received the Ph.D. degree in Telecommunications from the Signal Processing Lab at KTH Royal Institute of Technology, Stockholm, Sweden. He is currently working as a Post-Doc in the same lab.

His research interests include wireless multiantenna communications, resource allocation, feedback design, estimation theory, stochastic signal processing, and mathematical optimization.
For his work on optimization of multicell MIMO communications, he received a Best Paper Award at the 2009 International Conference on Wireless Communications \& Signal Processing (WCSP'09) and a Best Student Paper Award at the 2011 IEEE International Workshop on Computational Advances in Multi-Sensor Adaptive Processing (CAMSAP'11).
\end{IEEEbiography}

\begin{IEEEbiography}[{\includegraphics[width=1in,height=1.25in,clip,keepaspectratio]{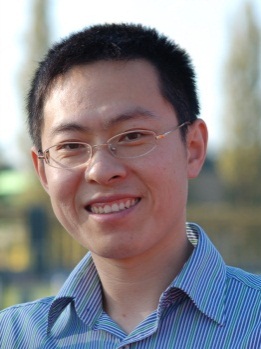}}]{Gan Zheng}
(S'05-M'09) received the BEng and MEng degrees from Tianjin University, China, in 2002 and 2004, respectively, both in Electronic and Information Engineering, and the Ph.D. degree in Electrical and Electronic Engineering from The University of Hong Kong, Hong Kong, in 2008.

He then worked as a Research Associate at University College London (UCL), London, UK. Since September 2010, he has been working as a Research Associate at the Interdisciplinary Centre for Security, Reliability and Trust (SnT), University of Luxembourg, Luxembourg. His research interests are in the general area of signal processing for wireless communications, with particular emphasis on multiuser multiple-input multiple-output (MIMO) system, cognitive and cooperative system, physical layer security and multibeam satellite communications.
 
Dr.~Zheng received the award for Researcher Exchange Programme from British Council to visit KTH Royal Institute of Technology in Sweden hosted by Professor Bj\"orn Ottersten, during September-November 2009. He received a Best Paper Award at the 2009 International Conference on Wireless Communications \& Signal Processing (WCSP'09) held in Nanjing, China.
\end{IEEEbiography}

\begin{IEEEbiography}[{\includegraphics[width=1in,height=1.25in,clip,keepaspectratio]{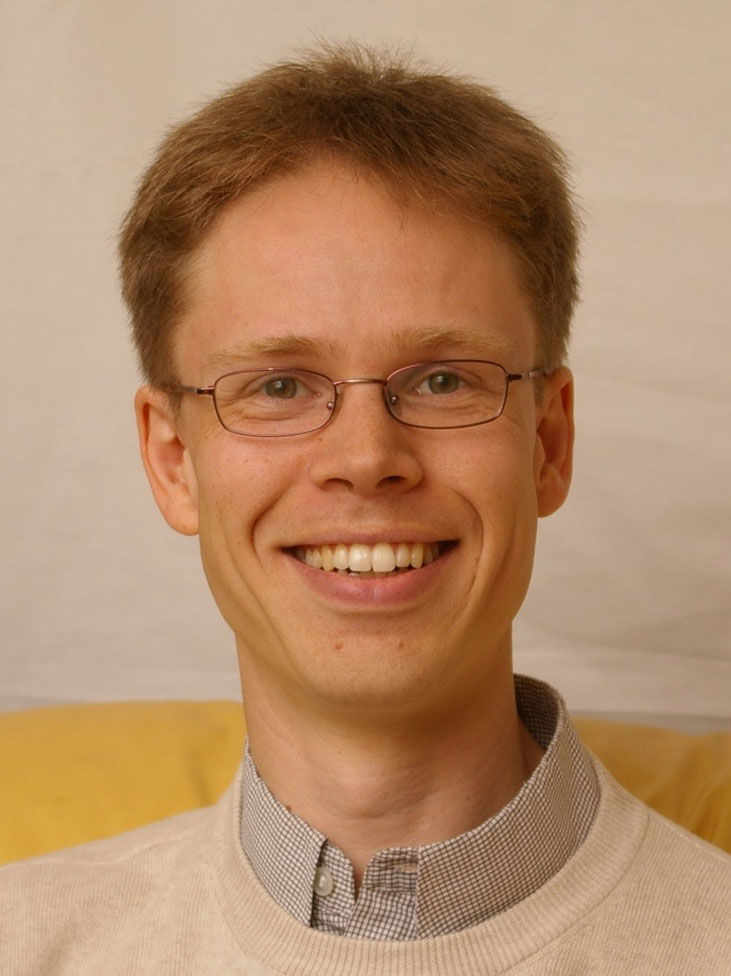}}]{Mats Bengtsson}
(M'00-SM'06) received the M.S. degree in computer science
from Link\"oping University, Link\"oping, Sweden, in 1991 and the Tech.~Lic.
and Ph.D. degrees in electrical engineering from the KTH Royal Institute of
Technology, Stockholm, Sweden, in 1997 and 2000, respectively.

From 1991 to 1995, he was with Ericsson Telecom AB Karlstad. He
currently holds a position as Associate Professor at the Signal
Processing Laboratory, School of Electrical Engineering, KTH. His
research interests include statistical signal processing and its
applications to communications, multi-antenna processing, radio resource
management, and propagation channel modelling. Dr. Bengtsson served as
Associate Editor for the IEEE TRANSACTIONS ON SIGNAL PROCESSING
2007-2009 and is a member of the IEEE SPCOM Technical Committee.
\end{IEEEbiography}

\begin{IEEEbiography}[{\includegraphics[width=1in,height=1.25in,clip,keepaspectratio]{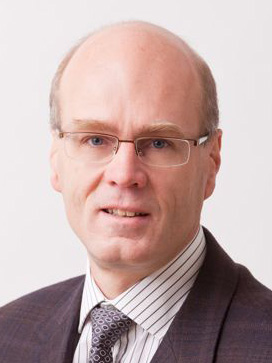}}]{Bj\"orn Ottersten}
(S'87-M'89-SM'99-F'04) was born in Stockholm, Sweden, in 1961. He
received the M.S. degree in electrical engineering and applied
physics from Link\"oping University, Link\"oping, Sweden, in 1986
and the Ph.D. degree in electrical engineering from Stanford
University, Stanford, CA, in 1989.

He has held research positions at the Department of Electrical
Engineering, Link\"oping University; the Information Systems
Laboratory, Stanford University; and the Katholieke Universiteit
Leuven, Leuven, Belgium. During 1996-1997, he was Director of
Research at ArrayComm Inc., San Jose, CA, a start-up company based
on Ottersten's patented technology. In 1991, he was appointed
Professor of Signal Processing at the KTH Royal Institute of
Technology, Stockholm, Sweden. From 2004 to 2008, he was Dean of the
School of Electrical Engineering at KTH, and from 1992 to 2004 he
was head of the Department for Signals, Sensors, and Systems at KTH.
He is also Director of security and trust at the University of
Luxembourg. His research interests include wireless communications,
stochastic signal processing, sensor array processing, and
time-series analysis.

Dr. Ottersten has coauthored papers that received an IEEE Signal
Processing Society Best Paper Award in 1993, 2001, and 2006. He has
served as Associate Editor for the IEEE TRANSACTIONS ON SIGNAL
PROCESSING and on the Editorial Board of the IEEE Signal Processing
Magazine. He is currently Editor-in-Chief of the EURASIP Signal
Processing Journal and a member of the Editorial Board of the
EURASIP Journal of Advances in Signal Processing. He is a Fellow of
IEEE and EURASIP. He is one of the first recipients of the European
Research Council advanced research grant.
\end{IEEEbiography}

\end{document}